\pdfoutput=1
\documentclass[12pt]{article}

\font\grande=cmr9.5 scaled \magstep4
\font\medio=cmr9.5 scaled \magstep2
\outer\def\beginsection#1\par{\medbreak\bigskip
      \message{#1}\leftline{\bf#1}\nobreak\medskip
\vskip-\parskip
      \noindent}
\usepackage{graphicx} 
\begin{document}
\bibliographystyle {unsrt}

\titlepage

\begin{flushright}
CERN-PH-TH/2015-196
\end{flushright}

\vspace{10mm}
\begin{center}
{\grande Statistical anisotropy from inflationary magnetogenesis}\\
\vspace{1.5cm}
 Massimo Giovannini
 \footnote{Electronic address: massimo.giovannini@cern.ch}\\
\vspace{1cm}
{{\sl Theory Department, CERN, 1211 Geneva 23, Switzerland }}\\
\vspace{0.5cm}
{{\sl INFN, Section of Milan-Bicocca, 20126 Milan, Italy}}
\vspace*{0.5cm}
\end{center}

\vskip 0.6cm
\centerline{\medio  Abstract}
Provided the quantum fluctuations are amplified 
in the presence of a classical gauge field configuration the resulting curvature 
perturbations exhibit a mild statistical anisotropy which should be sufficiently weak not to conflict with 
current observational data. The curvature power spectra induced by
weakly anisotropic initial states are computed here for the first time when the electric and the magnetic gauge couplings 
evolve at different rates as it happens, for instance, in the relativistic theory of van der Waals interactions. 
 After recovering the results valid for coincident gauge couplings, the constraints imposed by the isotropy and the homogeneity 
of the initial states are discussed. The obtained bounds turn out to be more stringent than naively expected and cannot be ignored 
when discussing the underlying magnetogenesis scenarios.
\vskip 0.5cm

\noindent

\vspace{5mm}

\vfill
\newpage
\renewcommand{\theequation}{1.\arabic{equation}}
\setcounter{equation}{0}
\section{Introduction}
\label{sec1}
Over the last decade the temperature and the polarization anisotropies of the Cosmic Microwave Background (CMB in what 
follows) have been scrutinized with the aim of finding specific hints signalling a minute breaking of rotational invariance of the 
power spectrum of curvature perturbations. Both the WMAP \cite{WMAP1} and the Planck experiments published dedicated 
analyses with the aim of  estimating the size of this admittedly small effect whose physical consequences 
could be potentially significant. In particular the seven and nine years WMAP \cite{wmap7a,wmap7b,wmap9} 
data and the two Planck \cite{PL2} releases specifically addressed this problem without reaching a conclusive evidence of the possible systematic 
nature of the effect whose statistical relevance is anyway not yet compelling.

In this situation various authors speculated that the anisotropic correction to the power spectrum of curvature perturbations could be the result 
of some form of inflationary dynamics leading to a perturbative breaking of rotational invariance 
(see \cite{anis1,anis2,anis3,anis4} for a time-ordered but still incomplete list of references). While different models 
have been examined, a plausible class of scenarios involves the presence of either electric or magnetic fields which must 
be sufficiently intense to affect the spectra of curvature perturbations but also extremely weak not to spoil the isotropy of the background. 
This possibility clashes, however, with a relatively well known obstruction represented by the 
so-called cosmic no-hair conjecture. In conventional inflationary models any finite portion of the universe gradually loses the memory of an initially 
imposed anisotropy or inhomogeneity so that the universe attains the observed regularity regardless of the initial boundary conditions \cite{hoyle1,hoyle2}. 

The electric or the magnetic energy densities
should be roughly constant for most of the inflationary evolution: this is the narrow path to obtain 
a sufficiently strong effect on the power spectrum and a comparatively negligible 
impact on the isotropy of the background geometry. In this respect a particularly plausible 
model is the one based on the coupling of the gauge kinetic term either to
the inflaton or to some other spectator field (see, for instance, \cite{DTA,DTB}).  
This scenario has been recently generalized to a class of models including, as a subcase, the relativistic theory of van der Waals interactions \cite{vdw0}.
This framework naturally leads to a different evolution of the electric and magnetic susceptibilities or, equivalently, 
of the electric and magnetic gauge couplings \cite{vdw}. In this paper we shall show that the possibility of achieving 
a substantial anisotropy in the power spectrum can be used to constrain the magnetogenesis scenarios based on the asymmetric 
evolution of the gauge couplings.
The plan of this paper is therefore the following. In section \ref{sec2} we shall discuss, in a unified manner, the magnetogenesis models 
based on the coupling of an Abelian gauge field to the inflaton or to some other spectator field. In section 3 we shall derive the evolution 
of curvature perturbations triggered by the presence of the gauge fields. The anisotropic power 
spectra and the constraints imposed on the whole scenario will be specifically derived in section 4. Section 5 contains the concluding remarks.

\renewcommand{\theequation}{2.\arabic{equation}}
\setcounter{equation}{0}
\section{Magnetogenesis and statistical anisotropy}
\label{sec2}
\subsection{General considerations}
We shall now consider a general form of the four-dimensional gauge action written in terms of two symmetric tensors (i.e. ${\mathcal M}_{\sigma}^{\rho}$ and  ${\mathcal N}_{\sigma}^{\rho}$) which may depend on the inflaton field $\varphi$  possibly supplemented by some spectator field $\sigma$:
\begin{equation}
S = -\frac{1}{16 \pi} \int d^4 x \, \sqrt{- G} \biggl[ \lambda(\varphi,\sigma) Y_{\alpha\beta} Y^{\alpha\beta} +  {\mathcal M}_{\sigma}^{\rho}(\varphi,\sigma) 
Y_{\rho\alpha}\, Y^{\sigma\alpha} - {\mathcal N}_{\sigma}^{\rho}(\varphi,\sigma) \widetilde{Y}_{\rho\alpha}\, \widetilde{Y}^{\sigma\alpha}
\biggr],
\label{action}
\end{equation}
where $G$ denotes the determinant of the four-dimensional metric $G_{\mu\nu}$; 
in Eq. (\ref{action}) $Y_{\mu\nu}$ and $\widetilde{Y}_{\mu\nu}$ are, respectively, the gauge field strength and its dual.
Conformally flat background geometries $G_{\mu\nu} = a^2(\tau) \eta_{\mu\nu}$ (where $\tau$ denotes the conformal 
time coordinate, $a(\tau)$ is the scale factor and $\eta_{\mu\nu}$ the Minkowski metric) will be the main focus of the present analysis but various 
considerations can also be applied to different backgrounds.

For specific choices of ${\mathcal M}_{\sigma}^{\rho}$ and  ${\mathcal N}_{\sigma}^{\rho}$, Eq. (\ref{action}) reproduces 
the relativistic theory of van der Waals interactions \cite{vdw0}. 
The detailed derivation of the equations of motion has been already discussed in Ref. \cite{vdw} together the relevant symmetries of the system. 
The evolution equations for the electric and magnetic fields 
shall then be written as\footnote{To derive Eqs. (\ref{first}), (\ref{second}) and (\ref{third}) we assumed ${\mathcal M}_{\alpha\beta} = \lambda_{E} u_{\alpha} u_{\beta}$ and 
${\mathcal N}_{\alpha\beta} = \lambda_{B} \overline{u}_{\alpha} \overline{u}_{\beta}$ where the generalized four-velocities 
are normalized gradients of the inflaton or of the spectator field \cite{vdw}. In this case  $\Lambda_{B}$ and $\Lambda_{E}$ are defined, respectively, as $\Lambda_{B} = \lambda + \lambda_{B}/2$
and  $\Lambda_{E} = \lambda + \lambda_{E}/2$. More general parametrizations 
of ${\mathcal M}_{\alpha\beta}$ and of ${\mathcal N}_{\alpha\beta}$ do change the explicit expressions of $\Lambda_{B}$ and $\Lambda_{E}$ in terms of the various couplings (e.g. 
$\lambda$, $\lambda_{B}$, $\lambda_{E}$ and possibly others) but do not affect the general form of the evolution equations (\ref{first}), (\ref{second}) and (\ref{third}).}:
\begin{eqnarray}
&& \vec{\nabla} \times \biggl( \sqrt{\Lambda_{B}} \vec{B} \biggr) = \partial_{\tau} \biggl( \sqrt{\Lambda_{E}} \vec{E} \biggr) + 4 \pi \vec{J},
\label{first}\\
&& \vec{\nabla} \times \biggl(\frac{\vec{E}}{\sqrt{\Lambda_{E}}}\biggr) + \partial_{\tau} \biggl(\frac{\vec{B}}{\sqrt{\Lambda_{B}}}\biggr) =0,
\label{second}\\
&& \vec{\nabla} \cdot \biggl(\frac{\vec{B}}{\sqrt{\Lambda_{B}}}\biggr)=0,\qquad \vec{\nabla}\cdot ( \sqrt{\Lambda_{E}}\, \vec{E} ) = 4 \pi \rho_{q},
\label{third}
\end{eqnarray}
where $\vec{J}$ and $\rho_{q}$ are the current and the charge densities. Note, furthermore, that in Eqs. (\ref{first}), (\ref{second}) and (\ref{third}) the 
electromagnetic fields\footnote{The explicit components of the fields strengths will be denoted, in what follows, as $Y_{i0} = a^2 e_{i}$ and $Y_{ij} = - a^2 \epsilon_{ijk} b^{k}$; in 
practice all the discussion will be conducted in terms of the rescaled fields $\vec{E}$ and $\vec{B}$. } have been rescaled through the electric and magnetic susceptibilities, i.e. 
$\vec{B} = a^2 \, \sqrt{\Lambda_{B}}\, \vec{b}$ and  $\vec{E} = a^2 \, \sqrt{\Lambda_{E}}\, \vec{e}$. 
Whenever $\vec{J} \to 0$ and $\rho_{q} \to 0$, Eqs. (\ref{first}), (\ref{second}) and (\ref{third}) are invariant under duality transformations generalizing the standard case \cite{duality1} of coincident gauge couplings.  

\subsection{Electric and magnetic gauge couplings}

Equations (\ref{first}), (\ref{second}) and (\ref{third}) can be expressed in terms of 
the electric and magnetic susceptibilities defined as  $\chi_{E} = \sqrt{\Lambda_{E}}$ and $\chi_{B} = \sqrt{\Lambda_{B}}$; the inverse 
of the susceptibilities are related to the 
corresponding gauge couplings as $g_{E} = \sqrt{4\pi/ \Lambda_{E}}$ and $g_{B} = \sqrt{4\pi/ \Lambda_{B}}$.
The time evolution of $g_{E}$ and $g_{B}$ during a quasi-de Sitter stage of expansion 
amplifies the gauge field fluctuations.  In this investigation the curvature perturbations induced by the amplified gauge field fluctuations 
will be used to constrain the rates of the evolution of the gauge couplings denoted hereunder by $F_{E}$ and $F_{B}$:
\begin{equation}
g_{E}(a) = \overline{g}_{E} \biggl(\frac{a}{a_{i}}\biggr)^{F_{E}}, \qquad g_{B}(a) = \overline{g}_{B} \biggl(\frac{a}{a_{i}}\biggr)^{F_{B}}, \qquad 
f = \frac{g_{B}^2}{g_{E}^2} = f_{i}  \biggl(\frac{a}{a_{i}}\biggr)^{2(F_{B}- F_{E})},
\label{gaugeC}
\end{equation}
where $a_{i}$ denotes the scale factor at the onset of the dynamical evolution of the gauge couplings and $f(a)$ measures 
the mismatch between $g_{E}$ and $g_{B}$.  The moment 
at which the largest wavelength of the curvature perturbations exits the Hubble radius (i.e. $a_{ex}$) 
may either be ${\mathcal O}(a_{i})$ or much larger than $a_{i}$. Even if the present considerations 
are general, for the specific discussions we shall preferentially consider the case where $a_{i}$ and $a_{ex}$ are of the same 
order but with $a_{i} < a_{ex}$.

The limit of exactly coincident coupling  corresponds to $f(a)=1$ during the whole evolution and, in this case, the standard results apply \cite{DTA,DTB}.  
Two extreme physical situations can be envisaged: the case when $f \to 1$ at the end of inflation (i.e. $f_{f} = {\mathcal O}(1)$, the couplings {\em converge} towards the end of inflation) and the case when $f\to 1$ at the beginning of inflation (i.e. $f_{i} = {\mathcal O}(1)$ the couplings converge at the onset of inflation but {\em diverge} at the and of inflation). These two limiting situations are purely illustrative and various intermediate possibilities are also physically plausible. Having said this, the constraints derived from  the impact of the gauge field fluctuations on the gauge-invariant curvature perturbations will be charted in the $(F_{B},\, F_{E})$ 
plane and the two benchmark cases of converging couplings (i.e. $f_{f} \to 1$) and of diverging couplings (i.e. $f_{i} \to 1$) will be specifically examined. 

If $F_{B}$ and $F_{E}$ are both positive the gauge couplings are 
initially small and get strong at the end of inflation. Conversely if $F_{B}$ and $F_{E}$ are both negative the gauge couplings may be strong at the beginning of inflation while they get weaker and weaker towards the end. This second situation seems to be the most natural in conventional inflationary models where, initially, the gravitational coupling is potentially very large during the pre-inflationary phase. In the class of models investigated in \cite{vdw} however, this choice is not 
mandatory\footnote{ In the case of coincident gauge couplings \cite{DTA,DTB}, a quasiflat magnetic field spectrum  realized in the case of a decreasing gauge coupling which gets progressively smaller during inflation. If the magnetic and the electric susceptibilities do not coincide \cite{vdw}, the allowed regions in the parameter space of inflationary magnetogenesis gets anyway wider in comparison with the conventional class of models where $F_{E} \to F_{B}$. }.

\subsection{Initial conditions and gauge field fluctuations}

The nature of the initial conditions for the evolution of the Abelian gauge fields depends  on the unknown features of the protoinflationary phase. For instance we could consider a globally neutral Lorentzian plasma as a possible remnant of a preinflationary stage of expansion and pose the problem of the suitable initial conditions for the evolution of the large-scale electromagnetic inhomogeneities. During the protoinflationary regime, the Weyl invariance of the Ohmic current guarantees that the comoving conductivity is approximately constant.  When the electric fields are negligible thanks to the large
conductivity of the protoinflationary plasma, the magnetic field is supported by a static solenoidal current obeying, from Eq. (\ref{first}), 
$\vec{\nabla} \times \biggl( \sqrt{\Lambda_{B}} \vec{B} \biggr) \simeq 4 \pi \vec{J}$. Since the plasma is globally neutral 
the charge density vanishes in Eq. (\ref{third}). These initial conditions are generally inhomogeneous 
but they do not induce specific anisotropies. This analysis, in the case of coincident gauge couplings, can be 
found in \cite{weyl}. 

Another example of inhomogenous initial conditions not inducing specific anisotropies are the quantum mechanical initial data. 
In this case both the current density and the charge density vanish in Eqs. (\ref{first}) and (\ref{third}).
Quantum mechanical initial data are justified in the case where, for instance, the duration of inflationary phase is extremely long 
(i.e. $a_{i} \ll a_{ex}$ in our notations). Purely quantum mechanical initial data have been discussed in a variety of situations \cite{DTA,DTB}
in the case of coincident gauge couplings and also in the situation parametrized by Eq. (\ref{gaugeC}) \cite{vdw}.
There is a third type of initial data that could be imposed namely the
weakly anisotropic initial data: they do not modify the isotropic evolution of the background but may contain either an electric or a magnetic field (or both).  We are considering here the situation where the electric and the magnetic fields are sufficiently small not to 
change the background geometry but large enough to affect the evolution of the curvature inhomogeneities. 

In the absence of sources the evolution of the electric and magnetic fields can be separated into a homogeneous part (i.e. $E^{(0)}_{i}(a) $ and 
$B^{(0)}_{i}(a)$) supplemented by a fully inhomogeneous contribution denoted, in real space,  by $E^{(1)}_{i}(\vec{x}, a)$ and $B^{(1)}_{i}(\vec{x}, a)$.
The evolution of the homogenous contribution in terms of the susceptibilities (or of the corresponding gauge couplings) 
can be derived from Eqs. (\ref{first}) and (\ref{second}) by neglecting all the spatial gradients and the sources. The result can be 
expressed as follows:
\begin{eqnarray}
E_{i}(\vec{x}, a) &=& E^{(0)}_{i}(a) +  E^{(1)}_{i}(\vec{x}, a), \qquad E^{(0)}_{i}(a) = \frac{E_{0}}{\sqrt{\Lambda_{E}(a)}}\,\, \hat{n}_{i},
\label{sol1}\\
B_{i}(\vec{x}, a) &=& B^{(0)}_{i}(a) +  B^{(1)}_{i}(\vec{x}, a), \qquad B^{(0)}_{i}(a) = B_{0}\sqrt{\Lambda_{B}(a)} \,\,\hat{m}_{i},
\label{sol2}
\end{eqnarray}
where $E_{0}$ and $B_{0}$ are space-time constants while $\hat{n}_{i}$ and $\hat{m}_{i}$ are unit vectors defining the direction 
of the homogeneous components.  

\subsection{The inhomogneous energy-momentum tensor}
With the same notation of Eqs. (\ref{sol1}) and (\ref{sol2}) the first and second-order fluctuations of the energy density are defined as
\begin{equation}
\delta \rho_{E} = \delta \rho_{E}^{(1)} + \delta\rho_{E}^{(2)} +\,..., \qquad \delta \rho_{B} = \delta \rho_{B}^{(1)} + \delta\rho_{B}^{(2)} + ...\,
\label{deco}
\end{equation}
where the ellipses stand for the higher order in the perturbative expansion. The fluctuations appearing in Eq. (\ref{deco}) 
can be directly expressed in terms of the gauge field fluctuations and they 
are\footnote{This result holds, strictly speaking, when ${\mathcal N}_{\alpha\beta} =0$. The general result
is discussed below in connection with Eq. (\ref{endens}).}
\begin{eqnarray}
\delta \rho_{E}^{(1)}  &=& \frac{1}{4 \pi a^4} E_{i}^{(0)} E_{i}^{(1)}, \qquad  \delta\rho_{E}^{(2)} =  \frac{1}{8 \pi a^4} E_{i}^{(1)} E_{i}^{(1)},
\label{deco1}\\
\delta \rho_{B}^{(1)}  &=& \frac{1}{4 \pi a^4} B_{i}^{(0)} B_{i}^{(1)}, \qquad  \delta\rho_{B}^{(2)} =  \frac{1}{8 \pi a^4} B_{i}^{(1)} B_{i}^{(1)}.
\label{deco2}
\end{eqnarray}
According to Eqs. (\ref{sol1}) and (\ref{sol2}) a particularly relevant case is the one where, up to numerical factors,  $\delta \rho_{E}^{(1)}$ and  $\delta \rho_{B}^{(1)} $ are proportional to $ \hat{n}\cdot \vec{E}^{(1)}$ and to $ \hat{m}\cdot \vec{B}^{(1)}$. 
This situation is realized when the time dependence of $\Lambda_{E}$ and $\Lambda_{B}$ exactly matches 
the dilution factors of the energy density:
\begin{equation}
\sqrt{\Lambda_{E}} \propto \biggl(\frac{a}{a_{i}}\biggr)^{-2}, \qquad \sqrt{\Lambda_{B}} \propto \biggl(\frac{a}{a_{i}}\biggr)^{2},
\end{equation}
guaranteeing that the corresponding energy densities are constant.
The same expansion can be obtained for the pressure and for the total anisotropic stresses. In particular we have that 
$\Pi_{ij} = \Pi_{ij}^{(1)} +  \Pi_{ij}^{(2)}$; note, however, that $\Pi_{ij}^{(1)} =0$ and the first contribution comes 
to second order in the amplitude of the electric and magnetic fields.  The terms containing the spatial 
gradients of the inflaton (like, for instance, $\vec{\nabla} \varphi \cdot( \vec{B} \times \vec{E})$) contribute only to the third order.

The effect of the amplified gauge field fluctuations on the curvature perturbations depend not only on the 
 energy density but also on the other components of the energy momentum tensor.
It is useful to write down, in this perspective,  the energy-momentum tensor of the electric and magnetic inhomogeneities in their 
full generality:
\begin{eqnarray}
{\mathcal T}_{\mu}^{\nu} &=& \frac{1}{4\pi}\, \biggl[ - {\mathcal S}_{\mu}^{\nu} + \frac{1}{4} \, {\mathcal S} \, \delta_{\mu}^{\nu}\biggr],
\label{ENM1}\\
{\mathcal S}_{\mu}^{\nu} &=& \lambda Y_{\alpha\mu} \, Y^{\alpha\nu} + \frac{1}{2}\biggl( {\mathcal M}^{\rho}_{\mu} \, Y_{\rho\alpha} \, Y^{\nu\alpha} + {\mathcal M}^{\rho}_{\sigma} \, Y_{\rho\mu} \,Y^{\sigma\nu} \biggr)
\nonumber\\
&-& \frac{1}{2} \biggl( {\mathcal N}^{\rho}_{\mu} \, \tilde{Y}_{\rho\alpha} \, \tilde{Y}^{\nu\alpha} + {\mathcal N}^{\rho}_{\sigma} \tilde{Y}_{\rho\mu}\, \tilde{Y}^{\sigma\nu} \biggr).
\label{ENM2}
\end{eqnarray}
From Eq. (\ref{ENM2}) with simple algebra the explicit components of Eq. (\ref{ENM1}) can be formally written as:
\begin{eqnarray}
{\mathcal T}_{0}^{0} &=& \delta\rho_{B} + \delta\rho_{E}, 
\label{t00}\\
{\mathcal T}_{i}^{j} &=& - (\delta p_{E} + \delta p_{B}) \delta_{i}^{j} + \Pi_{i}^{j}, 
\label{tij}\\
{\mathcal T}_{0}^{i} &=&  \frac{1}{4 \pi a^4} \biggl[ \sqrt{\frac{\Lambda_{E}}{\Lambda_{B}}}  + \sqrt{\frac{\Lambda_{B}}{\Lambda_{E}}}  - \frac{\overline{\Lambda}_{B}}{ \sqrt{\Lambda_{E} \, \Lambda_{B}}} \biggr](\vec{E} \times \vec{B})^{i},
\label{ti0}
\end{eqnarray}
where, recalling the remarks of Eqs. (\ref{first}), (\ref{second}) and (\ref{third}), $\overline{\Lambda}_{B} = (\lambda - \lambda_{B}/2)$.  The fluctuations of the energy density, of the pressure and the 
total anisotropic stress are given explicitly  by:
\begin{eqnarray}
&& \delta\rho_{B} = 3 \,\delta p_{B} = \frac{B^2}{8\pi a^4} \biggl(\frac{\overline{\Lambda}_{B}}{\Lambda_{B}} \biggr), \qquad  
 \delta\rho_{E} = 3 \,\delta p_{E} = \frac{E^2}{8 \pi a^4},
\label{endens}\\
&& \Pi_{i j} = \Pi_{ij}^{(E)} +  \Pi_{ij}^{(B)},
\label{anstress1}\\
&& \Pi_{ij}^{(E)} = \frac{1}{4\pi a^4} \biggl[ E_{i} E_{j} - \frac{E^2}{3}  \delta_{i j} \biggr], \qquad  
\Pi_{ij}^{(B)} = \frac{1}{4\pi a^4} \biggl[  B_{i} B_{j} - \frac{B^2}{3} \delta_{i j} \biggr] \biggl(\frac{\overline{\Lambda}_{B}}{\Lambda_{B}} \biggr).
\label{anstress2}
\end{eqnarray}
Whenever $\Lambda_{B} = \overline{\Lambda}_{B}$ we must have that $\lambda_{B} =0$. In this 
case ${\mathcal N}_{\rho}^{\sigma} =0 $ in the action of Eq. (\ref{action}). Note also that when $\Lambda_{B} = \overline{\Lambda}_{B}$ the prefactor in Eq. 
(\ref{ti0}) reduces to $\sqrt{\Lambda_{E}/\Lambda_{B}}$. 
In explicit models $\overline{\Lambda}_{B}/\Lambda_{B} \to 0$ at the beginning of inflation and goes to $1$ at the end of inflation; this 
effect  reduces the contribution of the magnetic field to the total energy density. For the sake of simplicity we shall 
analyze the simplest situation namely the one corresponding to the case $\lambda_{B} \to 0$. The case $\lambda_{B} \neq 0$ can be recovered, if needed, by redefining the relevant components of the energy-momentum tensor through the ratio  $\overline{\Lambda}_{B}/\Lambda_{B}$.

\renewcommand{\theequation}{3.\arabic{equation}}
\setcounter{equation}{0}
\section{Magnetized curvature perturbations}
\label{sec3}
The evolution of the magnetized scalar modes can be studied in terms of two well known but slightly different  variables denoted 
hereunder by ${\mathcal R}$ and $\zeta$ whose physical interpretation depends on the coordinate system: for instance ${\mathcal R}$ 
measures the curvature perturbations on comoving orthogonal hypersurfaces while $\zeta$ defines the curvature perturbations on uniform density hypersurfaces\footnote{On uniform curvature hypersurfaces (which will be the ones adopted hereunder in the uniform curvature gauge) $\zeta$ corresponds to the  total density contrast.}. Both variables are invariant under infinitesimal coordinate transformations as required in the context of the  Bardeen formalism \cite{bard1}. 
When spatial gradients can be neglected as it happens in the large-scale limit,
 $\zeta$ and ${\mathcal R}$ are {\em approximately} the same. This is why the second-order (decoupled) evolution 
equations obeyed by ${\mathcal R}$ and ${\zeta}$ are formally very different and lead to the same results only in the 
large-scale limit. With these caveats the evolution of the magnetized perturbations can be easily derived by selecting the 
hypersurfaces where the curvature is uniform: on these hypersurfaces the derivation of the evolution equation of ${\mathcal R}$ is easier. In this respect
a consistent choice is represented by the so-called uniform curvature gauge  \cite{hwna,hwnb,hwnc} which has been successfully exploited in related contexts. 
In what follows the evolution equations of the magnetized perturbations will be derived and solved.

\subsection{Uniform curvature gauge}
In the uniform curvature gauge the scalar fluctuations of the four-dimensional geometry are parametrized by two different functions 
describing the inhomogeneities in the $(00)$ and $(0i)$ entries of the perturbed metric \cite{hwna,hwnb,hwnc}:
\begin{equation}
 \delta_{s} G_{00} = 2 a^2 \phi , \qquad \delta_{\mathrm{s}} G_{ij} = 0, \qquad 
 \delta_{s} G_{0i} = - a^2  \partial_{i} \beta,
\label{UC1} 
\end{equation}
where $\delta_{s}$ denotes the scalar fluctuation of the corresponding quantity. The choice of Eq. (\ref{UC1}) 
completely fixes the coordinate system and guarantees the absence of spurious gauge modes. In the gauge  (\ref{UC1}), up to a background 
dependent coefficient, $\phi$ coincides with ${\mathcal R}$ while $\zeta$ is instead proportional to the total density contrast\footnote{As usual we shall denote with the prime a derivation with respect to the conformal time coordinate $\tau$ and, as usual, ${\mathcal H} = a^{\prime}/a$.}:
\begin{equation}
{\mathcal R} = - \frac{{\mathcal H}^2}{{\mathcal H}^2 - {\mathcal H}'} \, \phi, \qquad \zeta = \frac{\delta_{s}\rho_{t} + \delta\rho_{B} + \delta\rho_{E} }{3 ( p_{t} + \rho_{t})},
\label{G1}
\end{equation}
where $\rho_{t}$ and $p_{t}$ are the energy density and pressure of the background sources while $\delta_{s}\rho_{t}$ (and later on $\delta_{s} p_{t}$) denote the corresponding fluctuations. 

Using  Eqs. (\ref{t00}), (\ref{tij}) and (\ref{ti0})  giving the perturbed form of the gauge energy-momentum tensor, 
the $(00)$ and $(0i)$ components of the perturbed Einstein equations become\footnote{Equations (\ref{HG1}) and 
(\ref{HM2}) are commonly referred to as, respectively, the Hamiltonian and the momentum constraints.}:
\begin{eqnarray}
&&{\mathcal H} \nabla^2 \beta + 3 {\mathcal H}^2 \phi = - 4\pi G a^2 \biggl[\delta_{s}\rho_{\mathrm{t}} + \delta \rho_{\mathrm{B}} + \delta\rho_{\mathrm{E}} \biggr],
\label{HG1}\\
&& ({\mathcal H}' - {\mathcal H}^2) \nabla^2 \beta - {\mathcal H} \nabla^2 \phi = 
4\pi G a^2 \biggl[(p_{t}+ \rho_{t}) \theta_{t} + P \biggr],
\label{HM2}
\end{eqnarray}
where $P$ is the three-divergence of the Poynting vector appearing in Eq. (\ref{ti0}); $\delta_{s}\rho_{t}$ and $\theta_{t}$ denote, respectively, the fluctuations 
of the total energy density of the background and the three-divergence of the total velocity field. With the same notations 
the spatial components of the perturbed Einstein equations are:
\begin{eqnarray}
&&  ({\mathcal H}^2 + 2 {\mathcal H}') \phi + {\mathcal H} \phi'  = 4\pi G a^2\bigg[ \delta_{s} p_{t}  - \Pi_{\mathrm{E}} + \Pi_{\mathrm{B}}\biggr],
\label{sep3}\\
&& \nabla^2\beta' + 2 {\mathcal H} \nabla^2\beta + \nabla^2\phi = 12 \pi G a^2  \biggl(\Pi_{\mathrm{E}} + \Pi_{\mathrm{B}}\biggr),
\label{sep4}
\end{eqnarray}
where, as already mentioned,  $\delta_{s} p_{t}$ denotes the fluctuation of the total pressure while $\Pi_{E}$ and $\Pi_{B}$ are the scalar projections of the total anisotropic stress defined in the standard manner, i.e. $\nabla^2 \Pi_{\mathrm{B}} = \partial_{i} \partial_{j} \Pi^{ij}_{\mathrm{B}}$ and 
$\nabla^2 \Pi_{\mathrm{E}}= \partial_{i} \partial_{j} \Pi^{ij}_{\mathrm{E}}$.

\subsection{Adiabatic evolution of magnetized curvature perturbations}

The evolution of the curvature perturbations can be different depending on the background sources but in the present context 
we shall bound our attention on the most relevant case of a single inflaton field $\varphi$. In this case 
we have that, in the uniform curvature gauge, $\delta_{s} \rho_{t} \equiv \delta \rho_{\varphi}$, $\delta_{s} p_{t}\equiv \delta p_{\varphi}$ and 
$\theta_{t} \equiv\theta_{\varphi}$ where
 \begin{eqnarray}
&& \delta\rho_{\varphi} =  (- \phi {\varphi'}^2 + \chi_{\varphi}' \varphi')/a^2 + 
V_{\,,\varphi} \chi_{\varphi}, \qquad \delta p_{\varphi} - c_{\varphi}^2 \delta \rho_{\varphi} = \frac{V_{\,\,,\varphi}}{6\pi G \varphi'} \nabla^2 \beta,
\label{single1}\\
&& c_{\varphi}^2 = \frac{\partial p_{\varphi}}{\partial \rho_{\varphi}} = 1 + \frac{2 a^2 
V_{\,\,,\varphi}}{3 {\mathcal H} \varphi'}, \qquad \theta_{\varphi} = - \frac{\nabla^2 \chi_{\varphi}}{\varphi'} - \nabla^2 \beta,
\label{single3}
\end{eqnarray}
where $V(\varphi)$ is the inflaton potential, $\chi_{\varphi}$ is the inflaton fluctuation defined in the gauge (\ref{UC1}) and $V_{\,,\varphi}$ is the derivative 
of the inflaton potential with respect to $\varphi$.
In the single field case the constraint of Eq. (\ref{HM2}) together with the background equations implies that 
$\nabla^2 \phi = 4\pi G[ \varphi' \nabla^2 \chi_{\varphi}/{\mathcal H} - a^2 P/{\mathcal H}]$; the divergence of the Poynting 
vector can be neglected as in the case of coincident gauge couplings so that $\phi = 4\pi G \varphi' \chi_{\varphi}/{\mathcal H}$ \cite{fluc1}.

From Eqs. (\ref{G1}) and (\ref{HG1}) we can easily show that $\zeta = {\mathcal R} - {\mathcal H} \nabla^2 \beta/{[12 \pi G a^2 (p_{t} + \rho_{t}) ]}$
where $a^2 (p_{t} + \rho_{t})=  \varphi^{\prime\, 2}$ in the particularly relevant case where the background sources are represented by a single inflaton field $\varphi$. As a consequence of the previous relation the large-scale solutions of ${\mathcal R}$ coincide with the large-scale solutions of $\zeta$. This observation, however, {\em does not} imply that the second-order evolution equations of ${\mathcal R}$ and $\zeta$ coincide.  In the absence of magnetized contribution the evolution equation of ${\mathcal R}$ coincides 
with the canonical normal mode identified by Lukash \cite{lukash} when the source of the background is represented by a perfect relativistic fluid. The 
evolution of $\zeta$ in the presence of magnetized curvature perturbations has been discussed in \cite{fluc1} and subsequently employed by various authors.

To derive the decoupled evolution equation for ${\mathcal R}$ we can sum up Eq. (\ref{HG1}) (multiplied by $c_{\varphi}^2$) and Eq. (\ref{sep3}); after simple manipulations 
the following equation can be easily obtained:
\begin{eqnarray}
{\mathcal R}^{\prime} &=& \Sigma_{{\mathcal R}} + \frac{{\mathcal H}^2}{4 \pi G \varphi^{\prime 2} } \nabla^2 \beta,
\label{single4}\\
\Sigma_{{\mathcal R}} &=& \frac{\mathcal H a^2}{\varphi^{\prime 2}} \biggl[ 
\biggl(c_{\varphi}^2 - \frac{1}{3}\biggr) (\delta\rho_{\mathrm{B}} + \delta \rho_{\mathrm{E}}) + \Pi_{\mathrm{E}} + \Pi_{\mathrm{B}} \biggr].
\label{single5} 
\end{eqnarray}
By taking the first derivative of Eq. (\ref{single4}) and by using (\ref{single5})  Eq. (\ref{sep4}) to 
eliminate the time derivative of the Laplacian of $\beta$ the decoupled equation obeyed by ${\mathcal R}$ becomes\footnote{We are assuming, as natural, the background equations written in the form ${\mathcal H}^2 - {\mathcal H}^{\prime} = 4 \pi G {\varphi}^{\prime\, 2}$ and $3 {\mathcal H}^2 = 8 \pi G( {\varphi}^{\prime\, 2}/2 + V a^2)$.}
\begin{equation}
{\mathcal R}'' + 2 \frac{z_{\varphi}'}{z_{\varphi}} {\mathcal R}' -  \nabla^2 {\mathcal R} = \Sigma_{{\mathcal R}}' + 2 \frac{z_{\varphi}'}{z_{\varphi}} \Sigma_{{\mathcal R}} + 
\frac{3 a^4}{z_{\varphi}^2} (\Pi_{\mathrm{E}} + \Pi_{\mathrm{B}}),\qquad z_{\varphi} = a \varphi^{\prime}/{\mathcal H}.
\label{zt0}
\end{equation}
The term containing ${\mathcal R}^{\prime}$ at the left hand side of Eq. (\ref{zt0}) can be eliminated by defining $q = - z_{\varphi} \, {\mathcal R}$ and 
Eq. (\ref{zt0}) gets modified as:
\begin{equation}
q^{\prime\prime}  -  \nabla^2 q - \frac{z_{\varphi}^{\prime\prime}}{z_{\varphi}} q  = -\frac{1}{z_{\varphi}} \frac{\partial( z_{\varphi}^2\, \Sigma_{{\mathcal R}})}{\partial \tau} -  \frac{3 a^4}{z_{\varphi}} (\Pi_{\mathrm{E}} + \Pi_{\mathrm{B}}).
\label{zt0a}
\end{equation}
When the only source of inhomogeneities is an irrotational fluid, Eqs. (\ref{zt0}) and (\ref{zt0a}) keep almost the same form, with few changes: 
\begin{equation}
{\mathcal R}'' + 2 \frac{z_{t}'}{z_{t}} {\mathcal R}' - c_{\mathrm{st}}^2 \nabla^2 {\mathcal R} = \Sigma_{{\mathcal R}}' + 
2 \frac{z_{t}'}{z_{t}} \Sigma_{{\mathcal R}} + \frac{3 a^4}{z_{t}^2} (\Pi_{\mathrm{E}} + \Pi_{\mathrm{B}}),\qquad z_{\mathrm{t}} = (a^2 \sqrt{p_{\mathrm{t}} + \rho_{\mathrm{t}}})/({\mathcal H} c_{st}),
\label{zt1}
\end{equation}
where $c_{st}^2 =\partial p_{t}/\partial\rho_{t}$.  Except for the source term due to the inhomogeneities of the gauge fields
$z_{\mathrm{t}} \,{\mathcal R}$ defines, up to an irrelevant sign,  the normal mode of an irrotational and relativistic fluid discussed by Lukash \cite{lukash}; the subsequent analyses of Refs.  \cite{KS,chibisov} follow exactly the same tenets of Ref. \cite{lukash} but in the case of scalar field matter;  the normal modes of Refs. \cite{lukash,KS,chibisov}  coincide with the (rescaled) curvature perturbations on comoving orthogonal hypersurfaces \cite{bard1,br1}. 

So far only the adiabatic case has been treated but the presence of non-adiabatic fluctuations  
can be easily incorporated in Eqs. (\ref{zt0}) and (\ref{zt1}) (see, in particular, the second paper of Ref. \cite{fluc1} for the case of coincident gauge couplings). 
Indeed defining the non-adiabatic pressure fluctuation $\delta p_{nad} = \delta p_{\mathrm{t}} - c_{st}^2 \delta\rho_{\mathrm{t}}$, the derivation leading to Eqs. (\ref{zt0}) and (\ref{zt1}) can be swiftly generalized and the result is that Eq. (\ref{zt1}) still holds in the presence of non-adiabatic 
pressure fluctuations provided $\Sigma_{{\mathcal R}}$ is replaced by a slightly different source function denoted hereunder by $\overline{\Sigma}_{{\mathcal R}}$:
\begin{equation}
\Sigma_{{\mathcal R}} \to \overline{\Sigma}_{{\mathcal R}} = - \frac{{\mathcal H}\, \delta p_{nad}}{(p_{\mathrm{t}} + \rho_{\mathrm{t}})} + \Sigma_{{\mathcal R}}.
\label{NAD}
\end{equation}

A non-adiabatic pressure fluctuation develops, for instance, when the background contains two scalar fields (for example the inflaton $\varphi$ and a spectator field $\sigma$)
When the energy density of the inflaton dominates against the energy density of the spectator field $\sigma$ the evolution 
of curvature perturbations will still be given by Eq. (\ref{zt0}) where $\Sigma_{{\mathcal R}}$ is replaced by $\overline{\Sigma}_{{\mathcal R}}$ 
and $\delta p_{nad}$ is now given by $(\delta p_{\sigma} - c_{\varphi}^2 \delta \rho_{\sigma})$
where $\delta p_{\sigma}$ and $\delta \rho_{\sigma}$ are the pressure and the energy density fluctuations associated with the spectator field\footnote{
In the uniform curvature gauge the definition of $\delta p_{\sigma}$ and $\delta \rho_{\sigma}$ is similar to Eq. (\ref{single1}) but with 
$\varphi \to \sigma$ and with $\chi_{\varphi} \to \chi_{\sigma}$.}. Conversely, if the energy densities of the two 
fields are comparable the total curvature perturbation can be written, in the uniform curvature gauge as 
\begin{equation}
\nabla^2 {\mathcal R}= \frac{z_{\varphi}\, {\mathcal H} \varphi'}{{\varphi'}^2 + {\sigma'}^2} \nabla^2 {\mathcal R}_{\varphi}+ 
\frac{z_{\sigma}\, {\mathcal H} \sigma'}{{\varphi'}^2 + {\sigma'}^2}  \nabla^2 {\mathcal R}_{\sigma}+ \frac{{\mathcal H} a^2}{{\varphi'}^2 + {\sigma'}^2} P,
\label{cr2}
\end{equation} 
where, in the uniform curvature gauge, ${\mathcal R}_{\varphi} = - \chi_{\varphi}/z_{\varphi}$ and ${\mathcal R}_{\sigma} = - \chi_{\sigma}/z_{\sigma}$. In this 
case the evolution of the quasi-normal modes of the system (i.e. ${\mathcal R}_{\varphi}$ and ${\mathcal R}_{\sigma}$) is coupled and the relevant source terms 
can be deduced in full analogy with the discussion of the adiabatic case.  If taken into account the non-adiabatic component will lead to the kind of mixed initial conditions 
for CMB anisotropies often discussed in the literature \cite{corra,corrb} also in the presence of large-scale magnetic fields. 

\subsection{Large-scale solutions}
We shall now focus on the adiabatic case and assume a single field inflationary background. Equations (\ref{zt0}) and (\ref{zt0a}) can be easily 
solved for $a> a_{ex}$ in the regime where the Laplacians are negligible\footnote{For future convenience the 
integration variable appearing in Eq. (\ref{largescale}) coincides with scale factor. As previously mentioned in section \ref{sec2},  $a_{ex}$ denotes the moment at which the fluctuation with the largest wavelength exits the Hubble radius.}:
\begin{eqnarray}
{\mathcal R}(\vec{x},a) &=& {\mathcal R}_{ad}(\vec{x},a) + \int_{a_{ex}}^{a} \frac{b\,d b}{\varphi^{\prime 2}} \biggl[ \biggl( c_{\varphi}^2 - \frac{1}{3}\biggr) (\delta \rho_{B} + \delta \rho_{E})
+ \Pi_{B} + \Pi_{E}\biggr]_{b} 
\nonumber\\
&+& 3 \int_{a_{ex}}^{a} \biggl(\frac{{\mathcal H}}{\varphi^{\prime}}\biggr)^2 \frac{d b}{b^2} \int_{a_{ex}}^{b}(\Pi_{E} + \Pi_{B}) \frac{b^{\prime \,3}}{{\mathcal H}(b^{\prime})} d b^{\prime}, 
\label{largescale}\\
 {\mathcal R}_{ad}(\vec{x},a) &=& {\mathcal R}_{*}(\vec{x}) + \frac{{\mathcal R}^{\prime}(\vec{x}, a_{ex})- \Sigma_{{\mathcal R}}(\vec{x}, a_{ex})}{{\mathcal H}_{ex}} \biggl[ \biggl(\frac{a_{ex}}{a}\biggr)^3 -1 \biggr].
\label{adsol}
\end{eqnarray}
 The second term appearing at the right hand side of Eq. (\ref{adsol}) is negligible for $a> a_{ex}$ 
while ${\mathcal R}_{*}(\vec{x})$ is the (asymptotically constant) adiabatic solution. While Eqs. (\ref{largescale})--(\ref{adsol}) 
have been derived in real space, they can be easily Fourier transformed whenever needed. 

If the contribution of the anisotropic stress is negligible Eq. (\ref{largescale}) can be further simplified and the result is:
\begin{equation} 
{\mathcal R}(\vec{x}, a) = {\mathcal R}_{ad}(\vec{x}, a) + \frac{2}{3} \int_{a_{ex}}^{a} \frac{ b \, d b}{\varphi^{\prime\, 2}}\biggl[ 1 + \frac{b^2 V_{,\,\varphi}}{{\mathcal H} 
\varphi^{\prime}} \biggr] (\delta \rho_{B} + \delta \rho_{E}).
\label{int1}
\end{equation}
Furthermore, since the slow-roll approximation can be safely adopted for $a> a_{ex}$, Eq. (\ref{int1}) becomes: 
\begin{equation}
{\mathcal R}(\vec{x}, a) = {\mathcal R}_{ad}(\vec{x}, a) - 2 \int_{a_{ex}}^{a} \frac{ d b}{b \epsilon(b) V(b)}  (\delta \rho_{B} + \delta \rho_{E}).
\label{int2}
\end{equation}
The situation described by Eq. (\ref{int2}) is exactly the one relevant for the present discussion. Indeed, recalling Eqs. (\ref{deco1}) and (\ref{deco2}) and the 
related considerations, we have that the anisotropic stress vanishes to first-order so that Eq. (\ref{int2}) becomes\footnote{Equation (\ref{ANISONE}) holds also when $\epsilon$ is not strictly constant even if, for concrete applications, we shall bound 
the attention on the situation where the slow-roll parameters are constant, at least approximately. Notice that the $2\pi$ factor appearing in Eq. 
(\ref{ANISONE}) follows, ultimately, from the $1/(8 \pi)$ of the energy-momentum tensor of the gauge fields.}:
\begin{equation}
{\mathcal R}(a,\vec{x}) = {\mathcal R}_{ad}(\vec{x}) - \frac{1}{2\pi} \int_{a_{ex}}^{a} \frac{ d \ln{b}}{b^{4}\, \epsilon(b) \, V(b)} \biggl[ E_{i}^{(0)}(b) E_{i}^{(1)}(b, \vec{x}) +  B_{i}^{(0)}(b) B_{i}^{(1)}(b,\vec{x}) \biggr],
\label{ANISONE}
\end{equation}
where $\epsilon(b)$ denotes the slow-roll parameter.  

The result of Eq. (\ref{ANISONE}) has been obtained by neglecting the Laplacian in Eq. (\ref{zt0}) but it follows also from the general solution of Eqs. (\ref{zt0}) and (\ref{zt0a}) 
after integration by parts. Indeed, from Eqs. (\ref{zt0}) or (\ref{zt0a}) we have, in Fourier space, that\footnote{In Eq. (\ref{exr1})  the various functions appearing in the source term are evaluated in Fourier space.}:
\begin{equation}
{\mathcal R}(k,\tau) = {\mathcal R}_{ad}(k,\tau) + \int_{\tau_{*}}^{\tau} d\tau^{\prime} \, \frac{G^{(\varphi)}_{k}(\tau,\,\tau')}{z_{\varphi}(\tau) \, z_{\varphi}(\tau')} \biggl[\frac{\partial}{\partial \tau'} ( z_{\varphi}^2 \Sigma_{{\mathcal R}}) + 3 a^4 (\Pi_{E} + \Pi_{B})\biggl]_{\tau'},
\label{exr1}
\end{equation}
where ${\mathcal R}_{*}(k,\tau)$ denotes the solution of the homogeneous equation with the appropriate boundary conditions. Denoting with $F_{k}(\tau)$ and $F^{*}_{k}(\tau)$ the two independent solutions of the homogeneous equation obeyed by $z_{\varphi} {\mathcal R}$ (i.e. Eq. (\ref{zt0a})), the corresponding Green's function is:
\begin{equation}
G^{(\varphi)}_{k}(\tau,\tau') = \frac{F_{k}(\tau') \, F^{*}_{k}(\tau) -F_{k}(\tau) \, F^{*}_{k}(\tau') }{W(\tau')},
\label{GR1}
\end{equation}
where $W(\tau') =[ F^{\prime}_{k}(\tau')\,F^{*}_{k}(\tau') - F^{*\prime}_{k}(\tau') F_{k}(\tau')]$  is the Wronskian of the solutions. The explicit form of the mode function is\footnote{In Eq. (\ref{prox14})  $\eta$ and $\epsilon$ denote the standard slow-roll parameters in the case of single field inflationary backgrounds; furthermore the normalization is given by $|{\mathcal N}_{\varphi}| = \sqrt{\pi/2}$. }
\begin{equation}
 F_{k}(\tau) = \frac{{\mathcal N}_{\varphi}}{\sqrt{2 k}}\, \sqrt{- k \tau} H_{\widetilde{\mu}}^{(1)}(-k \tau), \qquad 
\widetilde{\mu} = \frac{3 + \epsilon + 2 \eta}{2 ( 1 -\epsilon)}.
\label{prox14}
\end{equation}
The expression of the Green's function depends on the index $\widetilde{\mu}$  of the corresponding Hankel functions. Since $\epsilon\ll 1$ and $\eta \ll 1$, the Bessel index
$\widetilde{\mu}$ can be expanded in powers of the slow roll parameters and $\widetilde{\mu} \simeq 3/2 + 2 \epsilon +\eta$ and $\tilde{\mu} = 3/2 +\epsilon$. Consequently, to leading order in the slow roll expansion $\widetilde{\mu}\simeq  3/2$ and, in this limit, the explicit expressions of $G^{(\varphi)}_{k}(\tau,\,\tau^{\prime})$ is
\begin{equation}
G^{(\varphi)}_{k}(\tau,\tau') = \frac{1}{k} \biggl\{ \frac{\tau' - \tau}{k \, \tau'\,\tau} \cos{[k (\tau'- \tau)]} - \biggl(\frac{1}{k^2 \tau' \tau} + 1\biggr) \sin{[k(\tau'-\tau)]}\biggr\}.
\label{GR}
\end{equation}
Equation (\ref{ANISONE}) follows then immediately from Eq. (\ref{exr1}) after one integration by parts. The essential 
result, in this respect, is the following: 
\begin{equation}
z_{\varphi}^2(\tau')\, \Sigma_{{\mathcal R}}(k,\tau') \frac{\partial}{\partial \tau'} \biggl[ \frac{G^{(\varphi)}_{k}(\tau,\tau')}{z_{\varphi}(\tau) z_{\varphi}(\tau')} \biggr]
\to \frac{\Sigma_{{\mathcal R}}(k,\tau') }{3 \tau^{\prime 2}} \frac{\partial}{\partial \tau'}[ \tau^{\prime\, 3} - \tau^3]  \equiv  \Sigma_{{\mathcal R}}(k,\tau'),
\label{GR2}
\end{equation}
where the limit has been taken for $\tau^{\prime} > \tau$ and $k \tau^{\prime} < 1$ (valid at large scales). In summary, these considerations demonstrate 
that Eq. (\ref{ANISONE}) can be safely used for the explicit analysis: it has been obtained by solving the exact evolution equation at large scales and it is correctly reproduced by taking the large-scale limit of the exact solution.

\subsection{Quantum mechanical considerations}
Since the effective action obeyed by the curvature perturbations is given by:
\begin{equation}
S_{{\mathcal R}} = \int d^3x\, d\tau \biggl\{ \frac{z_{\varphi}^2}{2} \biggl[ (\partial_{\tau} {\mathcal R})^2 - (\partial_{i} {\mathcal R})^2 \biggr] + \biggl[ \partial_{\tau}(z_{\varphi}^2 \Sigma_{{\mathcal R}})+ 3 a^4 (\Pi_{E} + \Pi_{B}) \biggr] {\mathcal R} \biggr\},
\label{acR1}
\end{equation}
the corresponding Hamiltonian is given by the sum of the free and of the interacting parts as $H_{{\mathcal R}}(\tau)= H_{0}(\tau)  +H_{I}(\tau)$ where 
$H_{0}(\tau)$  and $H_{I}(\tau)$ are:
\begin{eqnarray}
H_{0}(\tau) &=& \frac{1}{2} \int d^{3} x \biggl[ \frac{ \pi_{{\mathcal R}}^2}{ z_{\varphi}^2} + z_{\varphi}^2 (\partial_{i} {\mathcal R})^2 \biggr], \qquad  \pi_{{\mathcal R}} = z_{\varphi}^2\, \partial_{\tau} {\mathcal R},
\nonumber\\
H_{I}(\tau) &=& - \int d^{3} x [ \partial_{\tau}( z_{\varphi}^2 \Sigma_{{\mathcal R}}) + 3 a^4 (\Pi_{E} + \Pi_{B}) ] {\mathcal R}.
\label{acR3}
\end{eqnarray}
The normal modes  and the corresponding momenta can be promoted to quantum field operators , i.e. ${\mathcal R}\to \hat{{\mathcal R}}$ and 
$\pi_{{\mathcal R}}\to \hat{\pi}_{{\mathcal R}}$ obeying canonical commutation relations at equal time\footnote{Units $\hbar =1$ will be used throughout.}  $[ \hat{{\mathcal R}}(\vec{x},\tau), \pi_{{\mathcal R}}(\vec{y},\tau)] = i \delta^{(3)}(\vec{x} - \vec{y})$. The evolution equations obeyed 
by the field operators are  $\partial_{\tau} \hat{\pi}_{{\mathcal R}} = i [ \hat{H}_{{\mathcal R}}, \hat{\pi}_{{\mathcal R}}]$ 
and $\partial_{\tau} \hat{{\mathcal R}} = i [ \hat{H}_{{\mathcal R}}, \hat{{\mathcal R}}]$. It is easy to show that Eq. (\ref{zt0}) also holds 
for the corresponding field operator.  
In the absence of electromagnetic sources the operator corresponding to the adiabatic solution of Eqs. (\ref{adsol}) and (\ref{exr1}) is given by:
 \begin{equation}
\hat{{\mathcal R}}_{ad}(\vec{x}, \tau) = \frac{1}{(2 \pi)^{3/2} } \int d^{3} k \, \hat{{\mathcal R}}_{ad}(\vec{k}, \tau) e^{- i \vec{k} \cdot \vec{x}}, \qquad \hat{{\mathcal R}}(\vec{k}, \tau) = \frac{ F_{k}(\tau) 
\hat{a}_{\vec{k}} + F_{k}^{*} \hat{a}^{\dagger}_{- \vec{k}} }{z_{\varphi}}.
\label{rr}
\end{equation}
where $[\hat{a}_{\vec{q}}, \hat{a}^{\dagger}_{\vec{p}}] = \delta^{(3)}(\vec{q} - \vec{p})$ and the mode functions $F_{k}$ and $F_{k}^{*}$ appearing in Eq. (\ref{rr}) have been already introduced in Eqs. (\ref{GR1}) and (\ref{prox14}).  The connection bewteen the Green functions  discussed in Eqs. (\ref{GR})--(\ref{GR1})  and the quantum discussion follows from the commutator of the field operators in Fourier space at different times:
\begin{equation}
[ \hat{{\mathcal R}}_{ad}(\vec{q}, \tau_{1}), \, \hat{{\mathcal R}}_{ad}(\vec{p}, \tau_{2}) ] =- i \frac{G_{q}(\tau_{1}, \tau_{2})}{z_{\varphi}(\tau_{1}) \, z_{\varphi}(\tau_{2})} \delta^{(3)}(\vec{q} + \vec{p}).
\label{rr1}
\end{equation}

As a consequence of the previous discussion, Eq. (\ref{ANISONE}) holds also in quantum mechanical terms when the field fluctuations are replaced by quantum operators. More precisely we have that 
\begin{equation}
\hat{{\mathcal R}}(\vec{x},a) = \hat{{\mathcal R}}_{ad}(\vec{x},a) - \frac{1}{2\pi} \int_{a_{ex}}^{a} \frac{ d \ln{b}}{b^{4}\, \epsilon(b) \, V(b)} \biggl[ E_{i}^{(0)}(b) \hat{E}_{i}^{(1)}(\vec{x},b) +  B_{i}^{(0)}(b) \hat{B}_{i}^{(1)}(\vec{x},b) \biggr],
\label{ANISONEop}
\end{equation}
where operators corresponding to the electric and magnetic fields are instead given by: 
\begin{eqnarray}
&& \hat{B}^{(1)}_{i}(\vec{x}, \eta) = - \frac{i\, \epsilon_{m n i}}{(2\pi)^{3/2}\, \sqrt[4]{f(\eta)}}  \sum_{\alpha} \int d^{3} k k_{m} e^{(\alpha)}_{n} 
\biggl[ \overline{F}_{k}(\eta)\, \hat{a}_{\vec{k}, \alpha} e^{- i \vec{k} \cdot \vec{x}}  - \overline{F}_{k}^{*}(\eta) \hat{a}^{\dagger}_{\vec{k}, \alpha}
e^{ i \vec{k} \cdot \vec{x}}\biggr], 
\nonumber\\
&& \hat{E}^{(1)}_{i}(\vec{x}, \eta) = - \frac{1}{(2\pi)^{3/2}\, \sqrt[4]{f(\eta)}}  \sum_{\alpha} \int d^{3} k \,e^{(\alpha)}_{i} 
\biggl[ \overline{G}_{k}(\eta)  \hat{a}_{\vec{k}, \alpha} e^{- i \vec{k} \cdot \vec{x}}  + \overline{G}_{k}^{*}(\eta) \hat{a}^{\dagger}_{\vec{k}, \alpha}e^{ i \vec{k} \cdot \vec{x}} \biggr].
\label{exp3}
\end{eqnarray}
The time variable $\eta$ appearing in Eq. (\ref{exp3}) is related to 
the conformal time coordinate as $d\tau = \sqrt{f}\, d\eta$. Using this new time parametrization\footnote{The time variable $\eta$ cannot be confused with the slow-roll parameter since the two quantities do not appear in the same context} the mode functions appearing in Eq. (\ref{exp3}) obey the following simple equations: 
\begin{eqnarray}
\frac{d^{2} \overline{F}_{k}}{d\eta^2} + \biggl[k^2 -  \sqrt{g_{B} g_{E} }\biggl(\frac{1}{\sqrt{g_{B} g_{E}}}\biggl)^{\bullet\bullet} \biggr]  \overline{F}_{k}=0,\qquad 
 \frac{d^{2} \overline{G}_{k}}{d\eta^2} + \biggl[k^2 -  \frac{(\sqrt{g_{B} g_{E}})^{\bullet\bullet}}{\sqrt{g_{B} g_{E}}}\biggr]\overline{G}_{k}=0.
\label{bardec2}
\end{eqnarray}
The explicit solutions for $\overline{F}_{k}$ and $\overline{G}_{k}$ can be directly obtained by solving 
Eq. (\ref{bardec2}) in the $\eta$ parametrization and the result is:
\begin{eqnarray}
&& \overline{F}_{k}(\eta) = \frac{{\mathcal N}}{\sqrt{ 2 k}} \, \sqrt{- k \eta} \, H_{\sigma}^{(1)}(- k \eta),\qquad \sigma = \frac{1 - 2 F_{E}}{2( 1 + F_{B} - F_{E})},
\label{sol1m}\\
&& \overline{G}_{k}(\eta) = - {\mathcal N}\, \sqrt{\frac{k}{2}}\, \sqrt{- k \eta} \, H_{\sigma-1}^{(1)}(- k \eta),
\label{sol2m}
\end{eqnarray}
where $|{\mathcal N}|= \sqrt{\pi/2}$ where $H^{(1)}_{\alpha}(z)$ denotes, in general, the Hankel function of the first kind with 
index $\alpha$ and argument $z$. Having solved the mode functions in terms of $\eta$ it is always possible to go back 
to the conformal time coordinate $\tau$ or even to the scale facto itself as we shall show explicitly in the next section. 

To compute the anisotropic corrections to the adiabatic power spectrum it 
will then be necessary to evaluate the two-point function of $\hat{{\mathcal R}}$. From Eq. (\ref{ANISONE}) the anisotropic correction to the two-point 
function of $\hat{{\mathcal R}}$ is related to to the two-point functions of the gauge field 
fluctuations in terms of the mode functions, the Fourier components of the field operators $\hat{B}^{(1)}_{i}(\vec{x},\tau)$ and $\hat{E}^{(1)}_{i}(\vec{x},\tau)$ are respectively: 
\begin{eqnarray}
&& \hat{B}^{(1)}_{i}(\vec{q},\,\eta) = - \frac{i}{\sqrt[4]{f(\eta)}}\, \epsilon_{m n i}\sum_{\alpha}  \, e^{(\alpha)}_{n} \, q_{m} \bigl[ \hat{a}_{\vec{q},\alpha} \,\,\overline{F}_{q}(\eta) 
+ \hat{a}^{\dagger}_{-\vec{q},\alpha} \,\,\overline{F}^{*}_{q}(\eta)\bigr], 
\label{Fex1}\\
&& \hat{E}_{i}^{(1)}(\vec{q},\,\eta) = \frac{1}{\sqrt[4]{f(\eta)}}\, \sum_{\beta} e^{(\beta)}_{i} \bigl[ \hat{a}_{\vec{q},\beta} \,\,\overline{G}_{q}(\eta) 
+ \hat{a}^{\dagger}_{-\vec{q},\beta} \,\,\overline{G}^{*}_{q}(\eta)\bigr].
\label{Fex2}
\end{eqnarray}
Using Eqs. (\ref{Fex1}) and (\ref{Fex2}) the explicit correlation functions of the electric and magnetic fluctuations can be computed in terms 
of the corresponding mode functions, namely
\begin{eqnarray}
&& \langle \hat{B}^{(1)}_{i}(\vec{k},\eta)\, \hat{B}^{(1)}_{j}(\vec{p},\eta) \rangle = \frac{k^{2}\, |\overline{F}_{k}(\eta)|^2}{ \sqrt{f(\eta)}}  P_{ij}(k)\,\delta^{(3)}(\vec{k} + \vec{p}),
\label{cc1}\\
&& \langle \hat{E}^{(1)}_{i}(\vec{k},\eta)\, \hat{E}^{(1)}_{j}(\vec{p}, \eta) \rangle =  \frac{|\overline{G}_{k}(\eta)|^2}{ \sqrt{f(\eta)}} \,P_{ij}(k)\,\delta^{(3)}(\vec{k} + \vec{p}),
\label{cc2}
\end{eqnarray}
where $P_{ij}(k)=(\delta_{ij} - k_{i} k_{j}/k^2)$. 

Before analyzing the anisotropic corrections to the power spectrum of curvature perturbationswe mention  that the
the Hamiltonian of the gauge fields can be easily written by using the $\eta$ parametrization and the result is: 
\begin{eqnarray}
H_{{\mathcal A}}(\eta) &=& \frac{1}{2} \int d^3 x \biggl[ \vec{\Pi}^{2} + 2 \frac{(\sqrt{\chi_{E} \chi_{B}})^{\bullet}}{\sqrt{\chi_{E} \chi_{B}}}\vec{\Pi} \cdot \vec{{\mathcal A}} + 
 \partial_{i} \vec{{\mathcal A}} \cdot \partial^{i} \vec{{\mathcal A}}\biggr],
 \label{hamA}\\
 \vec{\Pi} &=& \partial_{\eta} \vec{{\mathcal A}} - \frac{(\sqrt{\chi_{E} \chi_{B}})^{\bullet}}{\sqrt{\chi_{E} \chi_{B}}}\vec{{\mathcal A}},
\label{canA}
\end{eqnarray}
where the overdot denotes a derivation with respect to $\eta$. 
Equation (\ref{hamA}) is written in the Coulomb gauge which is the appropriate gauge to use since it is invariant under 
the Weyl rescaling of the four-dimensional metric \cite{weyl}.  For notational convenience Eq. (\ref{hamA}) is written in terms 
of the susceptibilities $\chi_{E}$ and $\chi_{B}$ while the parameter space of the model is 
more easily discussed in terms of the corresponding gauge couplings already introduced in Eq. (\ref{gaugeC}). In terms 
of the $\vec{\mathcal A}$ the electric and the magnetic field are given, respectively, by $\vec{E}(\vec{x},\eta) = - \vec{\Pi}(\vec{x},\eta)/\sqrt[4]{f(\eta)}$ and 
$\vec{B}(\vec{x},\eta) = \vec{\nabla}\times [\vec{{\mathcal A}}(\vec{x},\eta)/\sqrt[4]{f(\eta)}]$. From these expression and form 
the decomposition in Fourier modes of $\hat{\mathcal A}_{i}(\vec{x},\eta)$, Eq. (\ref{exp3}) follows immediately.

\renewcommand{\theequation}{4.\arabic{equation}}
\setcounter{equation}{0}
\section{Anisotropic power spectra of curvature modes}
\label{sec4}

The two point function of curvature perturbations in Fourier space can be computed  from Eq. (\ref{ANISONEop}) after some lengthy but straightforward algebra\footnote{The same 
result can be obtained by using Eq. (\ref{ANISONE}) by specifying separately the two-point functions of the gauge fields in Fourier space.}. Thus, the two-point function 
 in Fourier space becomes:
\begin{equation}
\langle \hat{{\mathcal R}}(a,\vec{k}) \hat{{\mathcal R}}(a,\vec{q}) \rangle = \langle \hat{{\mathcal R}}_{ad}(a,\vec{k}) \hat{{\mathcal R}}_{ad}(a,\vec{q}) \rangle
+ \frac{1}{4\pi^2} \int_{a_{ex}}^{a} \frac{d b}{b^{5} \epsilon V}\,  \int_{a_{ex}}^{a} \frac{d c}{c^{5} \epsilon V} {\mathcal F}(\vec{q},\,\vec{k};\, b,\, c),
\label{corr}
\end{equation}
where ${\mathcal F}(\vec{q},\,\vec{k};\, b,\, c)$ is the sum of four different contributions:
\begin{eqnarray}
&& E_{i}^{(0)}(b) E_{j}^{(0)}(c) \langle \hat{E}^{(1)}_{i}(\vec{q}, b)\, \hat{E}^{(1)}_{j}(\vec{k}, c)  \rangle 
+ B_{i}^{(0)}(b) B_{j}^{(0)}(c) \langle \hat{B}^{(1)}_{i}(\vec{q}, b)\, \hat{B}^{(1)}_{j}(\vec{k}, c)  \rangle 
\nonumber\\
&&+ E_{i}^{(0)}(b)  B_{j}^{(0)}(c) \langle \hat{E}^{(1)}_{i}(\vec{q}, b)\, \hat{B}^{(1)}_{j}(\vec{k}, c)  \rangle 
+ B_{i}^{(0)}(b)  E_{j}^{(0)}(c) \langle \hat{B}^{(1)}_{i}(\vec{q}, b)\, \hat{E}^{(1)}_{j}(\vec{k}, c)  \rangle.
\label{corr2}
\end{eqnarray} 
Note that in Eq. (\ref{corr2}) $E_{i}^{(0)}$ and  $B_{j}^{(0)}$ (with the appropriate combinations of indices) have been defined in Eqs. (\ref{sol1})--(\ref{sol2}) while $\hat{E}^{(1)}_{i}$ and $\hat{B}^{(1)}_{j}$ have been introduced in Eqs. (\ref{Fex1}) and (\ref{Fex2}). The power spectrum of curvature perturbations is defined, within the 
present conventions, as $\langle \hat{{\mathcal R}}(a,\vec{k}) \hat{{\mathcal R}}(a,\vec{q}) \rangle = 2\pi^2 {\mathcal P}_{{\mathcal R}}(k,a)\delta^{(3)}(\vec{k} + \vec{q})/k^3$. Recalling therefore Eqs. (\ref{Fex1}) and (\ref{Fex2}) into Eq. (\ref{corr2}) we can compute the explicit form of the anisotropic 
correction:
\begin{eqnarray}
{\mathcal P}_{{\mathcal R}}(k,a) &=& {\mathcal P}_{ad}(k) + {\mathcal P}_{anis}(k,\,a),
\label{ANISTWO}\\
{\mathcal P}_{anis}(k,\,a) &=& [ 1 - (\hat{n}\cdot\hat{k})^2] {\mathcal I}_{E}^2 (k,\,a,\, a_{ex}) + [ 1 - (\hat{m}\cdot\hat{k})^2] {\mathcal I}_{B}^2 (k,\,a,\, a_{ex})
\nonumber\\
&+& 2 [(\hat{n}\times\hat{m})\cdot\hat{k}]\, {\mathcal I}_{E}(k,\,a,\, a_{ex}) \,{\mathcal I}_{B}(k,\,a,\, a_{ex}).
\label{EX1}
\end{eqnarray}
In Eq. (\ref{ANISTWO}) ${\mathcal P}_{ad}(k)$ denotes the adiabatic contribution
while the integrals ${\mathcal I}_{E}( k,\,a,\, a_{ex})$ and ${\mathcal I}_{B}( k,\,a,\, a_{ex})$ are given, respectively, by:
\begin{eqnarray}
{\mathcal I}_{E}(k,\, a,\, a_{ex}) &=& \frac{ E_{0}}{2 \pi}\,\int_{a_{ex}}^{a} \frac{d b}{b^3\, V(b) \epsilon(b)} \frac{\sqrt{P_{E}(k, b)}}{\sqrt{\Lambda_{E}(b)}}, 
\label{EX1a}\\
{\mathcal I}_{B}(k,\, a,\, a_{ex}) &=& \frac{B_{0}}{2\pi}\, \int_{a_{ex}}^{a} \frac{d b}{b^3\, V(b) \epsilon(b)}\sqrt{P_{B}(k, b)}\sqrt{\Lambda_{B}(b)}.
\label{EX1b}
\end{eqnarray}
Note that the $P_{B}(k, b)$ and $P_{E}(k, b)$ appearing in Eqs. (\ref{EX1a}) and (\ref{EX1b}) are the electric and the magnetic power 
spectra defined as:
\begin{eqnarray}
P_{B}(k,\eta) = \frac{k^{5}}{2\, a^4 \pi^2\, \, \sqrt{f(\eta)}} \, |\overline{F}_{k}(\eta)|^2, \qquad P_{E}(k,\eta) = \frac{k^3}{2\, a^4 \pi^2\, \sqrt{f(\eta)}} \,  |\overline{G}_{k}(\eta)|^2.
\label{PP}
\end{eqnarray}
In Eq. (\ref{PP}) the power spectra appear as a function of $\eta$ but to perform explicitly the integrals of Eqs. (\ref{EX1a}) and (\ref{EX1b}) we rather 
need the power spectra in terms of the corresponding scale factors. To comply with this statement the mode functions of  Eqs. (\ref{sol1m}) and (\ref{sol2m}) can be first expressed in the conformal time parametrization (by means of the definition $d \tau = \sqrt{f} \,d\eta$) and then rewritten as a function of the scale 
factor during the quasi-de Sitter stage of expansion.

As an interesting cross-check of the obtained results, we remark that  Eq. (\ref{EX1}) can also be obtained within the Schwinger-Keldysh approach
often dubbed as Òin-inÓ formalism (see, for instance, \cite{wein}). For this purpose we need to use 
the interaction Hamiltonian of Eq. (\ref{acR3}) and to recall that the connection between our Green function and and the commutator 
of two field operators at different times is given by Eq. (\ref{rr1}). The general expression of the $n$-point correlation function in the in-in formalism is given, for instance,  by Ref. \cite{wein} and it depends on an infinite sum over $N$: the $N=0$ term in  is simply the average of the product of the 
field operators in the interaction picture and gives the adiabatic tree-level adiabatic contribution, the $N=1$ term vanish, the $N=2$ gives the anisotropic power spectrum and so on and so forth for the higher orders. Each order contains the integrals of the average of commutators. For instance, in the case 
relevant to the present situation, we need to evaluate  $\langle [[ \hat{{\mathcal R}}(\vec{k}_{1}, \tau)\,  \hat{{\mathcal R}}(\vec{k}_{2}, \tau), H_{I}(\tau_{1})], H_{I}(\tau_2)]\rangle$ where the interaction Hamiltonian has been given in Eq. (\ref{acR3}). As shown in Eq. (\ref{rr1}), the commutator of the adiabatic solutions at different times gives the Green' s function (\ref{GR}) and this is the bridge between the two complementary approaches.

\subsection{Explicit form of the anisotropic contribution}
 The explicit expressions of the power spectra entering  Eqs. (\ref{EX1a}) and (\ref{EX1b}) and appearing in Eq. (\ref{PP}) 
 can be obtained by evaluating  the solutions of Eqs. (\ref{sol1m}) and (\ref{sol2m}) in the small argument limit of the corresponding Hankel functions \cite{abr1}.
The horizon crossing condition in terms of $\eta$ (i.e. $k \eta_{ex} = {\mathcal O}(1)$) {\em is not} equivalent to the 
standard condition implemented in the $\tau$ parametrization (i.e.  $k \tau_{ex} = {\mathcal O}(1)$). After some simple algebra we can therefore reobtain the magnetic power spectrum already derived in Ref. \cite{vdw}:
\begin{eqnarray}
P_{B}(k,\,b,\, \sigma,\mu) &=& H^4 \,\,{\mathcal Q}_{B}(\sigma,\mu) \, \, f(b)^{|\sigma| -1} \,  \,\biggl(\frac{k}{b H}\biggr)^{5 - 2 |\sigma|}, 
\label{PSB}\\
{\mathcal Q}_{B}(\sigma,\,\mu) &=& \frac{\Gamma^2(|\sigma|)}{\pi^3} \, 2^{ 2 |\sigma| -3} \, |1+ \mu|^{ 2 |\sigma|-1},
\nonumber
\end{eqnarray}
where $\sigma$ has been already introduced in Eq. (\ref{sol1m}) while $\mu$ measures the difference\footnote{In Eq. (\ref{prox14}) a variable called $\widetilde{\mu}$ has been introduced as the index of the Hankel function entering the 
adiabatic power spectrum. Clearly $\widetilde{\mu}$ and the $\mu$ variable of Eq. (\ref{defsigma}) are totally unrelated. Similar comment holds for $\sigma$ appearing in Eq. (\ref{defsigma}) and the notation employed in section \ref{sec2} for a generic spectator field: since the two quantities never appear in the same context 
there cannot be any confusion.} in the rate of evolution 
of the electric and magnetic gauge couplings of Eq. (\ref{gaugeC}):
\begin{equation}
\sigma = \frac{1 - 2 F_{E}}{2(1 + F_{B} - F_{E})}, \qquad \mu = \frac{F}{2} = F_{B} - F_{E}.
\label{defsigma}
\end{equation}
Similarly thanks to Eq. (\ref{sol2m}) the electric power spectrum is: 
 \begin{eqnarray}
P_{E}(k,\,b,\,\sigma,\mu) &=& H^4 \,\,{\mathcal Q}_{E}(\sigma,\mu)  \, \, f(b)^{|\sigma-1| -1} \,  \,\biggl(\frac{k}{b H}\biggr)^{5 - 2 |\sigma\, -\,1 |},
\label{PSE}\\
{\mathcal Q}_{E}(\sigma,\,\mu) &=&  \frac{\Gamma^2(|\sigma-1|)}{\pi^3} \, 2^{ 2 |\sigma-1| -3} \, |1+ \mu|^{ 2 |\sigma-1|-1}.
\nonumber
\end{eqnarray}
Note that in the plane $(F_{B},\, F_{E})$ there is a singular trajectory, namely $1 + F_{B} - F_{E} =0$ where 
 $\sigma$ diverges. This singularity is not physical and stems from the fact that for $F_{E} = F_{B} +1$ 
 the gauge couplings evolve exponentially in $\eta$.  

We now insert Eqs. (\ref{PSB}) and (\ref{PSE}) into Eqs. (\ref{EX1a}) and (\ref{EX1b}) and recall the relations 
of $\Lambda_{E}$ and $\Lambda_{B}$ to the gauge couplings (see Eq. (\ref{gaugeC})); 
the corresponding integrals can be performed in explicit terms and the result is:
\begin{eqnarray}
&& {\mathcal I}_{E}(k,\, a,\, a_{ex}) = \frac{\overline{E}_{0}\, H^2\, \sqrt{{\mathcal Q}_{E}} }{4 \pi^{3/2} \,V \,\epsilon\, \alpha_{E} } \, f_{ex}^{(|\sigma -1| -1)/2}
\biggl(\frac{k}{a_{ex} H}\biggr)^{\beta_{E}} \biggl[ \biggl(\frac{a}{a_{ex}} \biggr)^{ \alpha_{E}} -1 \biggr], 
\label{EX2a}\\
&& {\mathcal I}_{B}(k,\, a,\, a_{ex}) = \frac{\overline{B}_{0}\, H^2 \sqrt{{\mathcal Q}_{B}}}{\sqrt{\pi} V\,\epsilon\, \alpha_{B}} \,
\biggl(\frac{k}{a_{ex} H}\biggr)^{\beta_{B}} \biggl[ \biggl(\frac{a}{a_{ex}} \biggr)^{ \alpha_{B}} -1 \biggr],
\label{EX2b}
\end{eqnarray}
where we found convenient to redefine $E_{0}$ an $B_{0}$ by introducing  $\overline{E}_{0} = g_{E}(a_{ex}) E_{0}/a_{ex}^2$ and 
$\overline{B}_{0} = B_{0}/[a_{ex}^2 g_{B}(a_{ex})]$.
In Eqs. (\ref{EX2a}) and (\ref{EX2b})  $(\alpha_{E},\,\alpha_{B})$, $(\beta_{E},\, \beta_{B})$ and $({\mathcal Q}_{E},\, {\mathcal Q}_{B})$ are all functions 
of $F_{E}$ and $F_{B}$. In particular $(\alpha_{E},\, \beta_{E})$ and $(\alpha_{B},\, \beta_{B})$ are given by:
\begin{eqnarray}
\alpha_{E}(F_{E}, F_{B}) &=& (\mu + 1) |\sigma - 1| + F_{E} - \mu - 9/2, \quad \beta_{E}(F_{E}, F_{B}) = 5/2 - |\sigma -1|,
\label{AB1}\\
\alpha_{B}(F_{E}, F_{B}) &=& (\mu + 1) |\sigma | - F_{B} - \mu - 9/2, \quad \beta_{B}(F_{E}, F_{B}) = 5/2 - |\sigma|.
\label{AB2}
\end{eqnarray}
The variables $\alpha_{X}$ and $\beta_{X}$ (with $X= E,\, B$)  are solely functions of $F_{E}$ and $F_{B}$ since both 
$\sigma$ and $\mu$ only depend upon $(F_{E},\, F_{B})$ according to Eq. (\ref{defsigma}).
Equations (\ref{EX2a}) and (\ref{EX2b}) hold when  $\alpha_{E}\neq0$ and  $\alpha_{B}\neq 0$. If  
$\alpha_{E}=0$ and $\alpha_{B}= 0$, Eqs. (\ref{EX2a}) and (\ref{EX2b}) become, respectively, 
\begin{eqnarray}
{\mathcal I}_{E}(k,\, a,\, a_{ex}) &=& \frac{\overline{E}_{0}\, H^2\, \sqrt{{\mathcal Q}_{E}}}{4 \pi^{3/2} V\,\epsilon} \,
\biggl(\frac{k}{a_{ex} H}\biggr)^{\beta_{E}}\, \ln{(a/a_{ex})},  
\label{EX3a}\\
{\mathcal I}_{B}(k,\, a,\, a_{ex}) &=& \frac{\overline{B}_{0}\,H^2\, \sqrt{{\mathcal Q}_{B}}}{\sqrt{\pi}  V\,\epsilon} \,
\biggl(\frac{k}{a_{ex} H}\biggr)^{\beta_{B}}\,\ln{(a/a_{ex})}.
\label{EX3b}
\end{eqnarray}
If $\beta_{E} = \beta_{B} = 0$ in  Eqs. (\ref{EX3a}) and (\ref{EX3b}) the corrections to the power spectra are logarithmically sensitive to the 
duration of inflation (i.e. they just 
\subsection{Phenomenological considerations}
The total power spectrum of curvature perturbations in the presence of anisotropic contributions changes depending upon 
the specific initial conditions. In the case of electric initial conditions, for instance, we will have that the total power spectrum is\footnote{For reasons of opportunity related to the way the observational data are presented (see e.g. \cite{wmap7b,wmap9,anis1}) the total power spectra of curvature perturbations  have been pametrized as in Eq. (\ref{electric0}). This parametization 
corresponds to the one of Ref. \cite{anis1} with the difference that, in the present case, the factor $g_{*}^{(X)}$ (with $X= E,\, B,\, BE$) can also 
depend $k$.}:
\begin{eqnarray}
{\mathcal P}_{{\mathcal R}}(k) &=& {\mathcal A}_{{\mathcal R}} \biggl(\frac{k}{k_{p}}\biggr)^{n_{s} -1} \biggl[ 1 + g_{*}^{(E)} (\hat{k}\cdot\hat{n})^2\biggr],
\qquad g_{*}^{(E)} = - \frac{4 {\mathcal Q}_{E} \overline{\Omega}_{E} {\mathcal G}_{E}}{3 \epsilon +4 {\mathcal Q}_{E} \overline{\Omega}_{E} {\mathcal G}_{E}},
\nonumber\\
{\mathcal G}_{E} &=& \frac{1}{\alpha_{E}^2} \biggl(\frac{k}{a_{ex} H}\biggr)^{2 \beta_{E}} f_{ex}^{|\sigma -1| -1} \biggl[ \biggl(\frac{a}{a_{ex}}\biggr)^{\alpha_{E}} -1\biggr]^2,
\label{electric0}
\end{eqnarray}
where $\overline{\Omega}_{E} = \overline{E}_{0}^2/(8\pi V)$. The explicit expression of ${\mathcal Q}_{E}$ has been already given in Eq. (\ref{PSE}).
In the case of magnetic initial conditions Eq. (\ref{electric0})  is replaced by
\begin{eqnarray}
{\mathcal P}_{{\mathcal R}}(k)&=& {\mathcal A}_{{\mathcal R}} \biggl(\frac{k}{k_{p}}\biggr)^{n_{s} -1} \biggl[ 1 + g_{*}^{(B)} (\hat{k}\cdot\hat{m})^2\biggr],
\qquad g_{*}^{(B)} = - \frac{64 \pi^2 {\mathcal Q}_{B} \overline{\Omega}_{B} {\mathcal G}_{B}}{3 \epsilon + 64 \pi^2 {\mathcal Q}_{B} \overline{\Omega}_{B} {\mathcal G}_{B}},
\nonumber\\
{\mathcal G}_{B} &=& \frac{1}{\alpha_{B}^2} \biggl(\frac{k}{a_{ex} H}\biggr)^{2 \beta_{B}} f_{ex}^{|\sigma -1| -1} \biggl[ \biggl(\frac{a}{a_{ex}}\biggr)^{\alpha_{B}} -1\biggr]^2, 
\label{magnetic0}
\end{eqnarray}
where $\overline{\Omega}_{B} = \overline{B}_{0}^2/(8\pi V)$. Note, as in the case of ${\mathcal Q}_{E}$ that the explicit expression of ${\mathcal Q}_{B}$ has been
already given in Eq. (\ref{PSB}). If the electric and magnetic fields
are simultaneously present we can have also mixed initial data:
\begin{eqnarray}
{\mathcal P}_{{\mathcal R}}(k) &=& {\mathcal A}_{{\mathcal R}} \biggl(\frac{k}{k_{p}}\biggr)^{n_{s} -1} \biggl\{ 1 + g_{*}^{(BE)} [(\hat{n}\times\hat{m})\cdot\hat{k}]^2\biggr\},
\nonumber\\
 g_{*}^{(BE)} &=& - \frac{16 \pi \sqrt{{\mathcal Q}_{B} {\mathcal Q}_{E}} \sqrt{\overline{\Omega}_{B} \overline{\Omega}_{E}} {\mathcal G}_{BE}}{3 \epsilon +16 \pi \sqrt{{\mathcal Q}_{B} {\mathcal Q}_{E}} \sqrt{\overline{\Omega}_{B} \overline{\Omega}_{E}} {\mathcal G}_{BE}},
\nonumber\\
{\mathcal G}_{BE} &=& \frac{1}{\alpha_{B} \alpha_{E}} \biggl(\frac{k}{a_{ex} H}\biggr)^{\beta_{E}+ \beta_{B}} f_{ex}^{\frac{|\sigma| + |\sigma -1|}{2} -1} \biggl[ \biggl(\frac{a}{a_{ex}}\biggr)^{\alpha_{B}} -1\biggr] \biggl[ \biggl(\frac{a}{a_{ex}}\biggr)^{\alpha_{E}} -1\biggr].
\label{mixed0}
\end{eqnarray}
The anisotropic corrections to the power spectrum of curvature perturbations have been derived under the hypothesis that the electric and the 
magnetic fields have a negligible impact on the evolution equations of the background geometry. Thus Eqs. (\ref{electric0}), (\ref{magnetic0}) 
and (\ref{mixed0}) are valid provided $\overline{\Omega}_{E} \ll 1$ and $\overline{\Omega}_{B} \ll 1$; this means, in practice, 
that Eqs. (\ref{electric0}), (\ref{magnetic0}) and (\ref{mixed0}) can be approximated as: 
\begin{eqnarray}
g_{*}^{(E)} &\simeq& - \frac{4}{3\epsilon}  \,\,{\mathcal Q}_{E} \,\overline{\Omega}_{E} \,\, {\mathcal G}_{E},\qquad 
g_{*}^{(B)} \simeq - \frac{64 \pi^2}{ 3 \epsilon} \,\, {\mathcal Q}_{B} \overline{\Omega}_{B} \,\, {\mathcal G}_{B},
\label{elmag}\\
g_{*}^{(BE)} &\simeq& - \frac{16 \pi}{3\epsilon} \,\,\sqrt{{\mathcal Q}_{B}\, {\mathcal Q}_{E}} \,\,\sqrt{\overline{\Omega}_{B} \,\overline{\Omega}_{E}} \,\,{\mathcal G}_{BE}.
\label{mixed1}
\end{eqnarray}
The argument pursued in the remaining part of this section is, in short, the following.
If the gauge couplings are coincident the flat spectrum of magnetic perturbations is realized when $F_{E} = F_{B} = -2$ and, in this case, 
the bounds stemming from the isotropy of the power spectra depend logarithmically on the duration of the inflationary phase.
If $F_{E} \neq F_{B}$  the flat magnetic power spectrum can also be obtained when $F_{E} \to (5 F_{B} + 4)/3$, as it follows from Eq. (\ref{PSB}) by setting $|\sigma| = 5/2$.
When $F_{E} \neq F_{B}$ the bounds stemming from the contribution of the gauge fluctuations to the curvature perturbation 
may show exponential sensitivity to the total number of inflationary efolds and this is why the curvature bounds are potentially more relevant 
in the $F_{E} \neq F_{B}$ case. In the next subsection we shall examine 
the bounds logarithmically dependent on the duration of inflation. In the remaining two subsections we shall discuss, respectively,  the bounds that are independent 
of the number of efolds and the bounds depending exponentially on the number of efolds. We shall finally draw the relevant exclusion 
plots in the $(F_{E},\,F_{B})$ plane and get to our conclusions.

\subsubsection{Bounds logartithmically dependent on the duration of inflation}

When the gauge couplings coincide  we have that $\mu \to 0$ and $F_{B} = F_{E} = F_{*}$. In this case from Eqs. (\ref{AB1}) 
and (\ref{AB2}) we have 
\begin{eqnarray}
\alpha_{E} &=& F_{*} - 9/2 + \frac{|1 + 2 F_{*}|}{2}, \qquad \beta_{E} = 5/2 -  |1 + 2 F_{*}|/2,
\label{AB1a}\\
\alpha_{B} &=& | 1 - 2 F_{*}|/2 - 9/2 - F_{*}, \qquad \beta_{B} = 5/2 - | 1 - 2 F_{*}|/2.
\label{AB2a}
\end{eqnarray}
Two particularly significant cases are the magnetic initial conditions (i.e. $\overline{E}_{0} =0$) with $\beta_{B} =0$ 
and the electric initial conditions (i.e. $\overline{B}_{0} =0$) with $\beta_{E} =0$. In these two cases both $g_{*}^{(E)}$ and 
$g_{*}^{(B)}$ are independent on the wavenumber and they are given by:
\begin{equation}
g_{*}^{(E)} \simeq - \frac{ 3 \overline{\Omega}_{E}}{\epsilon \pi^2} N_{ex}^2, \qquad 
g_{*}^{(B)} \simeq - \frac{ 48 \overline{\Omega}_{B}}{\epsilon} N_{ex}^2.
\label{AB1b}
\end{equation}
For the benchmark values $N_{ex} = {\mathcal O}(65)$ and $\epsilon = {\mathcal O}(10^{-2})$ we have that $\overline{\Omega}_{B}$ 
(or $\overline{\Omega}_{E}$) must be 
${\mathcal O}(10^{-9})$ (or smaller) if we want the anisotropic contribution to the power spectrum to be ${\mathcal O}(0.1)$ (or smaller).
This result agrees with the figures already obtained in the literature (see e.g. last two papers of Ref. \cite{anis4}).
When the gauge couplings coincide and the anisotropy parameters are scale-invariant there are two possible situations: 
either the magnetic power spectrum is also scale invariant (and the electric power spectrum is violet) or the electric power 
spectrum is scale invariant (and the magnetic power spectrum is red). The case of scale-invariant magnetic power 
spectrum is phenomenologically viable since magnetic fields ${\mathcal O}(10^{-2})\,\,\mathrm{nG}^2$ can be safely 
produced \cite{vdw} at the onset of galactic rotation\footnote{The power spectra of the electric and magnetic fluctuations 
have the dimensions of energy densities so that they are correctly measured in $nG^2$ ($ 1\, \mathrm{nG} = 10^{-9} \mathrm{G}$). 
Furthermore, in the present terminology, violet and red spectra are, respectively,  steeply increasing and decreasing as a function of the comoving wavenumber.}. 
The case of electric initial conditions supplemented by a scale-invariant electric power spectrum
is instead not phenomenologically viable \cite{vdw}. 
In summary we can say that, in the case of coincident gauge couplings, no 
further constraints on the  model itself can stem form the analysis of curvature perturbations. 

\subsubsection{Bounds independent on the number of efolds}

Since the induced curvature anisotropy must be negligible all over the dynamical evolution,  it should also be subleading, in particular,  few efolds 
after the given wavelength exceeded the Hubble radius. Therefore
the following bounds must hold for the electric and magnetic initial conditions: 
\begin{eqnarray}
&& \frac{4 Q_{E}}{ 3 \epsilon \alpha_{E}^2} \Omega_{E} f_{ex}^{3/2} < {\mathcal O}(0.1),\qquad \frac{64 \pi^2}{3 \epsilon \alpha_{B}^2} Q_{B} \Omega_{B} f_{ex}^{3/2} < {\mathcal O}(0.1),
\label{boun12}\\
&&\frac{16 \pi}{3 \epsilon \alpha_{B} \alpha_{E}} \sqrt{Q_{B} Q_{E}} \sqrt{\Omega_{E} \Omega_{B}} f_{ex}^{3/2} < {\mathcal O}(0.1),
\label{boun3}
\end{eqnarray}
where we consider the experimental upper limits on the anisotropic contribution to be at most ${\mathcal O}(0.1)$ \cite{wmap7a,wmap7b,wmap9,PL2}.
Equations (\ref{boun12}) and (\ref{boun3}) are obtained by evaluating the anisotropy for $a > a_{ex}$ but $a = {\mathcal O}(a_{ex})$. These 
relations are easily derived by recalling that $k\eta_{ex} = {\mathcal O}(1)$ implies also $k/(a_{ex} H) = {\mathcal O}(\sqrt{f_{ex}})$ 
since $d\tau = \sqrt{f} \,d\eta$. The two complementary cases of 
 diverging gauge couplings (i.e. $f(a_{i}) = f_{i} = {\mathcal O}(1)$ ) and of converging gauge couplings 
 (i.e. $f(a_{f})=f_{f} = {\mathcal O}(1)$) are not exhaustive but  they can be used to illustrate the nature of the bounds.
 
Consider, for the sake of simplicity, the case $f_{i} =1$ as illustrative of the case of diverging gauge couplings. In this 
case $f_{ex} \simeq (a_{ex}/a_{i})^{2 (F_{B} - F_{E})}$. As long as the relevant modes exit few efolds after the onset of inflation 
a potentially large term can be easily compensated by the relative smallness of $\overline{\Omega}_{E}$. 
Conversely in the case $f_{f} =1$ we have
$f_{ex} \simeq (a_{ex}/a_{f})^{2 (F_{B} - F_{E})}$. But this is nothing but $\exp{[ - 2 N_{ex} (F_{B} - F_{E})]}$ where $N_{ex}$ denotes 
the number of efolds elapsed since $a_{ex}$.
This number is pretty small iff $F_{E} < F_{B}$ but it is very large otherwise.  To ensure the validity of the constraints 
of Eqs. (\ref{boun12}) and (\ref{boun3}) we therefore have to demand $F_{E} < F_{B}$.  
The case $f_{f} = {\mathcal O}(1)$ is not exactly independent on the number of efolds and it is partly similar to the bounds derived in the
following subsection.

In summary we can say that the case of diverging gauge couplings is not constrained at $k \eta_{ex} ={\mathcal O}(1)$ while 
the case of converging gauge couplings is strongly constrained and, to be conservative, we should demand 
$F_{E} < F_{B}$. In this case the region of the parameter space is drastically reduced.  Apparently, a  
way out would be to postulate that $\overline{\Omega}_{E} = \overline{\Omega}_{B} =0$: this would mean that the case $f_{f} = {\mathcal O}(1)$ 
is incompatible with the presence of an initial electric or magnetic field. This way out is simplistic: to second order 
the contribution of the electric and magnetic fields to the power spectra will present the same problem. The second-order 
contribution does not produce the dependence on a specific direction and arises even if the initial state is only the vacuum 
\cite{fluc1}. In this case the analysis valid for coincident gauge couplings can be easily extended and the supplementary 
contribution to the power spectrum of curvature perturbations will depend on $f_{ex}^{3}$ (rather  than $f_{ex}^{3/2}$ as in the 
present case). In conclusion the derived bound is genuinely physical and cannot be artificially ignored.
\begin{figure}[!ht]
\centering
\includegraphics[height=8cm]{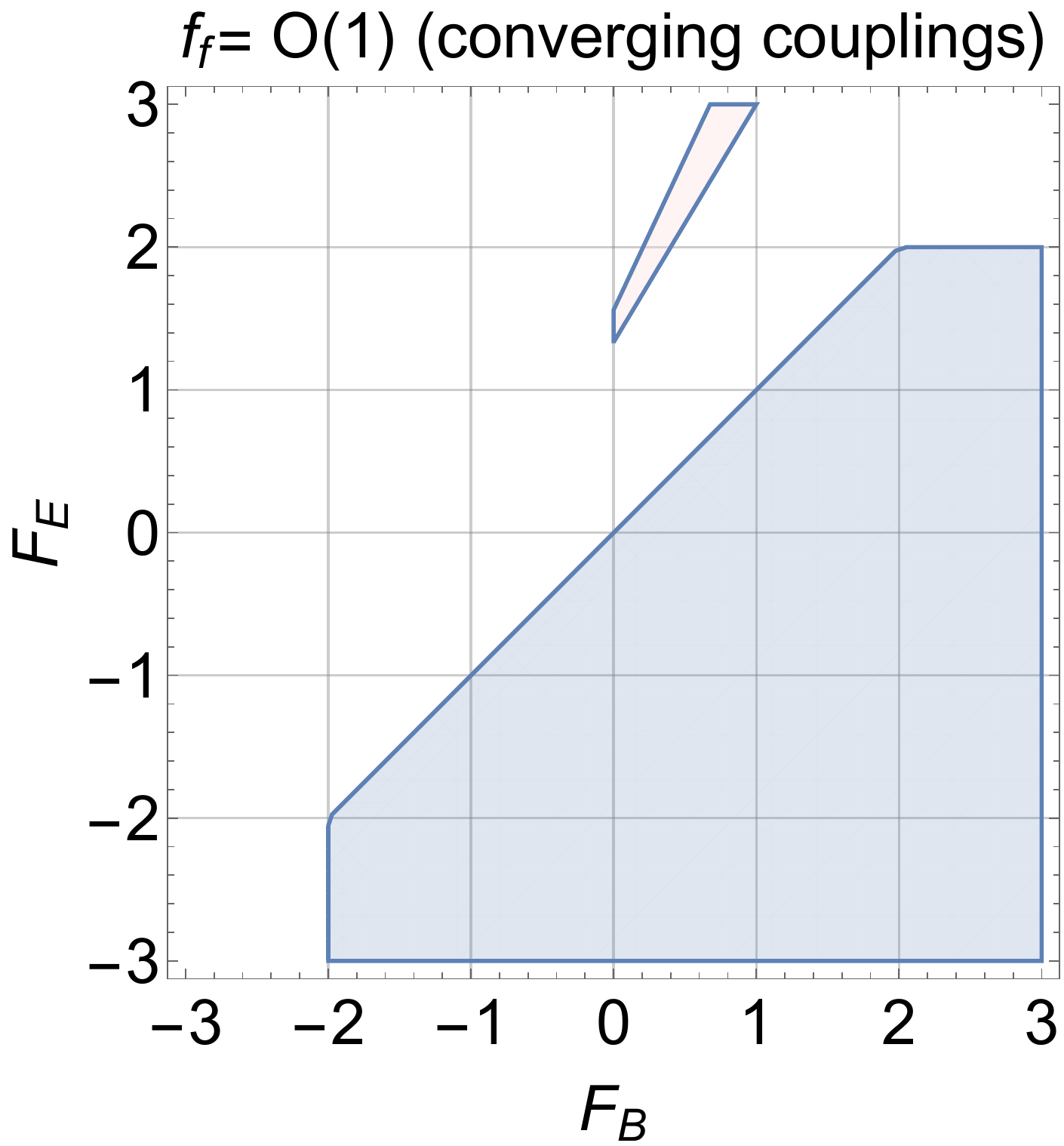}
\includegraphics[height=8cm]{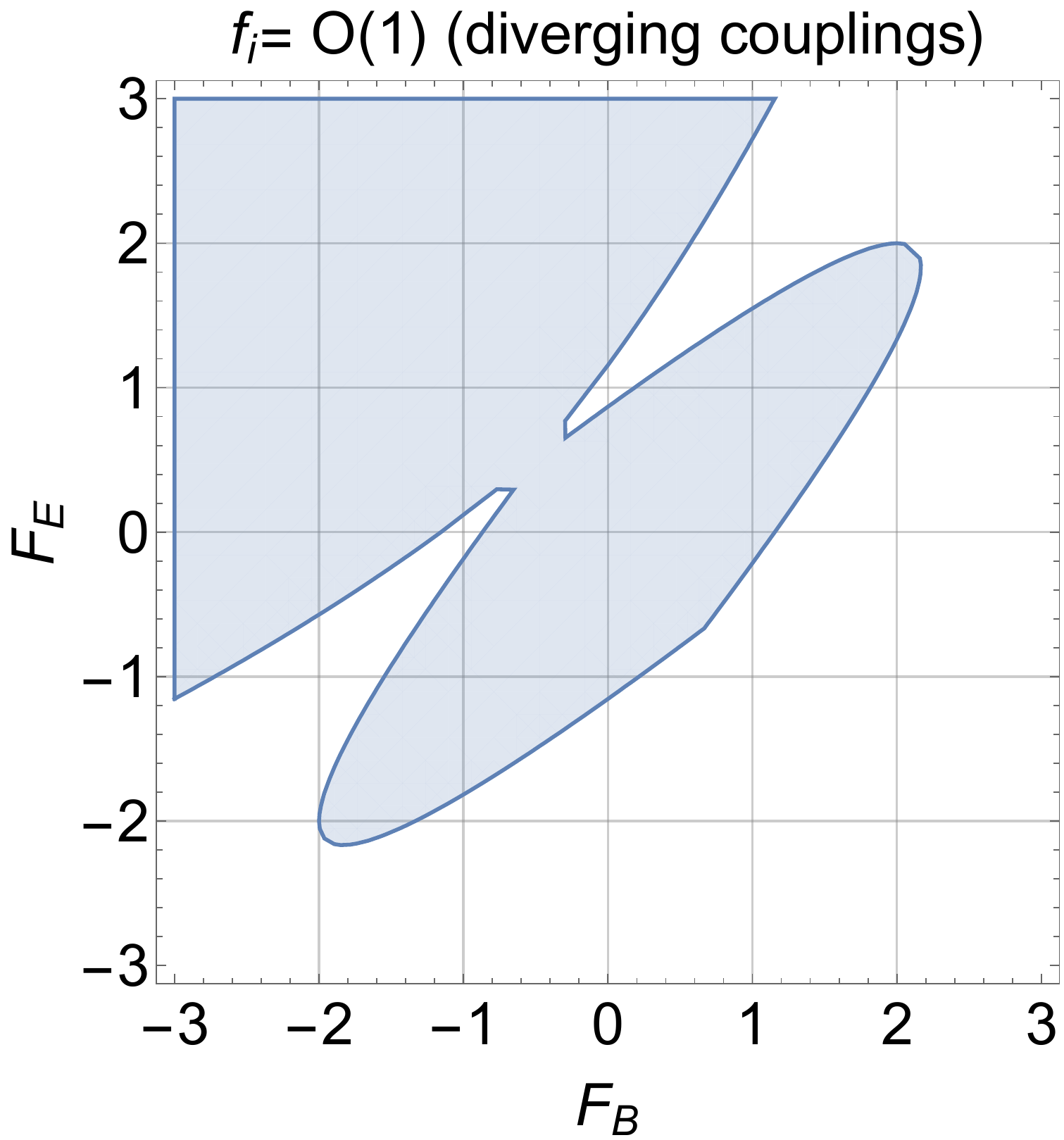}
\caption[a]{In both plots the shaded area  illustrates the allowed region of the parameter space. The left and right plots describe, respectively,
the case of converging and diverging gauge couplings.}
\label{Figure1}      
\end{figure}

\subsubsection{Bounds exponentially dependent on the number of efolds}

The bounds depending exponentially on the number of efolds can be obtained by evaluating the 
functions ${\mathcal G}_{E}(k,a)$, ${\mathcal G}_{B}(k,a)$ and ${\mathcal G}_{EB}(k,a)$ for $a = {\mathcal O}(a_{f})$ and by 
demanding that their relative contribution does not exceed the observational limits, in particular, at the maximal 
wavenumber of the spectrum. The general expressions of ${\mathcal G}_{E}(k,a)$, ${\mathcal G}_{B}(k,a)$ and ${\mathcal G}_{EB}(k,a)$
can be found, respectively, in Eqs. (\ref{electric0}), (\ref{magnetic0}) and (\ref{mixed0}). Since the same argument 
can be repeated for these three distinct functions we shall discuss analytically only  ${\mathcal G}_{E}(k_{max}, a_{f})$ and then
mention the results for the remaining cases. In the discussion we shall also assume\footnote{When $\alpha_{E}\to 0$ and $\alpha_{B}\to 0$ we showed that ${\mathcal G}_{E}$ and ${\mathcal G}_{B}$ depend 
logarithmically on the duration of the inflationary phase.} that $\alpha_{E} \neq 0$ and $\alpha_{B} \neq 0$.

From Eq. (\ref{electric0}) we can easily deduce the following expression:
\begin{eqnarray}
{\mathcal G}_{E}(k_{max}, a_{f}) = \frac{1}{\alpha_{E}^2} \biggl(\frac{k_{max}}{a_{f} H}\biggr)^{ 2 \beta_{E}} \biggl(\frac{a_{f}}{a_{ex}}\biggr)^{ 2 \alpha_{E} - 2 \mu [ |\sigma -1| -1] + 2 \beta_{E}}.
\label{firstb}
\end{eqnarray}
Since, by definition, $k_{max} \eta_{max} = {\mathcal O}(1)$ we must also have 
$k_{\max} \tau_{f} = {\mathcal O}(\sqrt{f_{f}})$. As before the case of converging and diverging gauge couplings can be treated separately.
In particular, if $f_{f} = {\mathcal O}(1)$, Eq. (\ref{firstb}) implies that 
the contribution of ${\mathcal G}_{E}(k_{max}, a_{f})$ will not explode iff:
\begin{equation}
\alpha_{E} -  \mu [ |\sigma -1| -1] +  \beta_{E} <0.
\label{firstc}
\end{equation}
But recalling the explicit values of $\alpha_{E}$, $\beta_{E}$, $\mu$ and $\sigma$ the last condition simply means that $F_{E} <2$.
The same argument leading to Eqs. (\ref{firstb}) and (\ref{firstc}) can be repeated in the case of Eqs. (\ref{magnetic0}) and (\ref{mixed0}).  
The analog of Eq. (\ref{firstc}) but derived from Eqs. (\ref{magnetic0}) and (\ref{mixed0}) will be, respectively,
\begin{equation}
\beta_{B} +  \alpha_{B} -  \mu[ |\sigma| -1] <0, \qquad \beta_{E} + \beta_{B} + \alpha_{E} + \alpha_{B} - \mu [ |\sigma| + |\sigma -1| - 2] <0. 
\label{secondc} 
\end{equation}
In more explicit term the two conditions of Eq. (\ref{secondc})  imply respectively $F_{B} +2 >0$ and $F_{E} - F_{B} <4$.

In the case of converging couplings the constraints obtained in the present and in the previous subsections are illustrated in Fig. \ref{Figure1}. The large shaded area extending through the fourth 
quadrant of the $(F_{E}, \, F_{B})$ plane represents the allowed region of the parameter space where 
all the constraints are safely satisfied. This region is bounded by the lines $F_{E} =2$, $F_{B} = -2$ and $F_{E} = F_{B}$.
The smaller region appearing in the first quadrant illustrates, as an example, a class of magnetogenesis models based on the case of converging 
gauge couplings \cite{vdw}: this area corresponds to the region $ 5 F_{B}/3 + 4/3 \leq F_{E} \leq 1.56 + 2.13 F_{B}$
and it is excluded since it does not overlap with the wider region allowed by the constraints on the isotropy 
of the power spectrum.  There are other regions in the first quadrant which are not excluded, including the frontier 
$F_{E} = F_{B}$. It is however clear that the allowed region extends more towards the fourth quadrant. This means, in practice 
that the models where $F_{B} >0$ and $F_{E}< 0$ are comparatively less constrained. 

Let us finally move to the case of diverging gauge couplings and assume for concreteness $f_{i} ={\mathcal O}(f_{ex}) = {\mathcal O}(1)$.
In this case the analog of Eq. (\ref{firstb}) becomes 
\begin{eqnarray}
{\mathcal G}_{E}(k_{max}, a_{f}) = \frac{1}{\alpha_{E}^2} f_{ex}^{  \beta_{E}} \biggl(\frac{a_{f}}{a_{ex}}\biggr)^{2 \mu \beta_{E} + 2 \alpha_{E} - 2 \mu [ |\sigma -1| -1] + 2 \beta_{E}},
\label{firstd}
\end{eqnarray}
implying that the contribution to the anisotropy is small for $f_{ex} = {\mathcal O}(1)$ provided
\begin{equation}
\beta_{E} (\mu+1) +  \alpha_{E} -  \mu [ |\sigma -1| -1] <0.
\label{secondd}
\end{equation}
The same argument can be applied to ${\mathcal G}_{B}(k_{max}, af)$ and ${\mathcal G}_{EB}(k_{max}, a_f)$. The results are, respectively,
\begin{equation} 
 (\mu + 1) \beta_{B} + \alpha_{B} - \mu [ |\sigma|-1] <0,\qquad
 (\beta_{E}+ \beta_{B}) (\mu + 1) + \alpha_{E} + \alpha_{B} - \mu [|\sigma| + |\sigma -1| -2 ] <0.
\label{thirdd}
\end{equation}
In the right plot of Fig. \ref{Figure1} the bounds obtained in the present and in the previous subsections 
are illustrated in the case of diverging gauge couplings. The various absolute values appearing in Eqs. (\ref{secondd}) and (\ref{thirdd}) 
make the frontier of the allowed region less intuitive. It is however clear that the least constrained portion of the parameter space 
is the second quadrant of the $(F_{E}, \, F_{B})$ plane where $F_{B}<0$ but $F_{E}$ is positive. 
As before there exist limited regions where both rates have the same sign. 

\renewcommand{\theequation}{5.\arabic{equation}}
\setcounter{equation}{0}
\section{Concluding remarks}
\label{sec5}
A generalized class of magnetogenesis scenarios based on the relativistic theory of 
van der Waals interactions implies an asymmetric evolution of the magnetic and electric gauge couplings. 
As the quantum fluctuations of the gauge fields are amplified, they also gravitate and even if they do not affect 
the evolution of the background itself, they contribute to the curvature power spectra which 
have been specifically computed in this paper during a quasi-de Sitter stage of expansion.

Depending on the sensitivity of the derived spectra to the total number of inflationary efolds three different classes of constraints may arise:
bounds logarithmically sensitive to the duration of inflation, bounds independent on the duration of inflation and finally 
 bounds which are exponentially sensitive to the number of inflationary  efolds.
In each of these cases the gauge couplings may either converge towards the end of inflation 
or diverge from the initial state. 
If the gauge couplings are converging they are of the same order at the end of 
inflation and, in this case, the allowed region corresponds, in practice, to the fourth 
quadrant in the $(F_{B},\,F_{E})$ plane where $F_{B}$ and $F_{E}$ are the rates of the evolution
of the gauge couplings in units of the Hubble rate.
If the gauge couplings are diverging they are of the same order at the onset of inflation but they can be
very different later on. In this case, except for few slices of the parameter space the allowed region 
falls almost entirely within the second quadrant of  the $(F_{B},\,F_{E})$ plane. It is relevant to stress that the scope of the obtained
constraints is exactly to pin down the regions of the parameter space where al the potentially large corrections to the curvature power spectra are negligible. 
In this sense the duration of inflation is immaterial for the allowed regions of the parameter space. 

The obtained results clearly show that the constraints point at the case where one of the two 
gauge couplings contracts and the other expands. On this basis, various classes 
of magnetogenesis scenarios can be excluded. There remains trajectories in the $(F_{B},\,F_{E})$ plane where 
the rates can be simultaneously negative or positive (like in the case  when $F_{E} \to F_{B}$) but these typically coincide with the boundaries of the allowed region. 
When the rates have the same sign the gauge couplings may converge at the end of inflation but these  
models lead to strong anisotropic corrections to the curvature power spectra and seem therefore excluded by the present 
conclusions. In a complementary perspective the obtained result might also suggest that there exist viable models of magnetogenesis 
based on the asymmetric evolution of gauge couplings but admitting a strongly anisotropic initial 
state which becomes isotropic at a later stage. This analysis is beyond the scopes of the present discussion. 

\section*{Acknowledgments}

The author wishes to thank T. Basaglia and J. Jerdelet and S. Rohr of the CERN scientific information service for their kind assistance.

\end{document}